\documentclass{jfm_old}

\usepackage{graphicx}
\usepackage{newtxtext}
\usepackage{newtxmath}
\usepackage{natbib}
\usepackage{hyperref}
\usepackage{caption3}
\usepackage{caption} 
\usepackage{subcaption}
\usepackage{xcolor}
\usepackage{contour}
\usepackage{color}

\newcommand{\MDGrevise}[1]{\textcolor{blue}{#1}}
\definecolor{ao(english)}{rgb}{0.0, 0.5, 0.0}

\title{Dynamics of a data-driven low-dimensional model of turbulent minimal Couette flow}

\author{Alec J. Linot\aff{1} \and Michael D. Graham\aff{1}\corresp{\email{mdgraham@wisc.edu}}}
\affiliation{\aff{1}Department of Chemical and Biological Engineering, University of Wisconsin-Madison, Madison WI 53706, USA}
\date{\today}

\begin{document}
\maketitle

\begin{abstract}
%

{
Because the Navier-Stokes equations are dissipative, the long-time dynamics of a flow in state space are expected to collapse onto a manifold whose dimension may be much lower than the dimension required for a resolved simulation. On this manifold, the state of the system can be exactly described in a coordinate system parameterizing the manifold. Describing the system in this low-dimensional coordinate system allows for much faster simulations and analysis. We show, for turbulent Couette flow, that this description of the dynamics is possible using a data-driven manifold dynamics modeling method. This approach consists of an autoencoder to find a low-dimensional manifold coordinate system and a set of ordinary differential equations defined by a neural network. Specifically, we apply this method to minimal flow unit turbulent plane Couette flow at $\Rey=400$, where a fully resolved solutions requires $\mathcal{O}(10^5)$ degrees of freedom. Using only data from this simulation we build models with fewer than $20$ degrees of freedom that quantitatively capture key characteristics of the flow, including the streak breakdown and regeneration cycle. At short-times, the models track the true trajectory for multiple Lyapunov times, and, at long-times, the models capture the Reynolds stress and the energy balance. For comparison, we show that the models outperform POD-Galerkin models with $\sim$2000 degrees of freedom. Finally, we compute unstable periodic orbits from the models. Many of these closely resemble previously computed orbits for the full system; additionally, we find nine orbits that correspond to previously unknown solutions in the full system.}

\end{abstract}

\section{Introduction} \label{sec:Intro}

A major challenge in dealing with chaotic fluid flows, whether it be performing experiments, running simulations, or interpreting the results, is the high-dimensional nature of the state. Even for simulations in the smallest domains that sustain turbulence (a minimal flow unit (MFU)){,} the state dimension may be $\mathcal{O}(10^5)$ \citep{Jimenez1991,Hamilton1995}. However, despite this nominal high-dimensionality, the dissipative nature of turbulent flows leads to the expectation that long-time dynamics {collapse} onto an invariant manifold of much lower dimension than the ambient dimension \citep{Hopf1948a}. By modeling the dynamics in a manifold coordinate system{,} simulations could be performed with a drastically lower{-dimensional} state representation{,} significantly speeding up {computations}. {Additionally, such a low-dimensional state representation is highly useful for downstream tasks like control or design.} Finding a low-dimensional -- or ideally a minimal dimensional -- parameterization of the manifold and an evolution equation for this parameterization are both challenges{. In this work we aim to address these challenges with a data-driven model, specifically for the task of reconstructing turbulent plane Couette flow.}


The classic way to perform dimension reduction {from data} is to use the proper orthogonal decomposition (POD), also known by {Principal Component Analysis (PCA) or} Karhunen-Lo\` eve decomposition \citep{Holmes1998}. This is a linear dimension reduction technique in which the state is projected onto the set of orthogonal modes that capture the maximum variance in the data. The POD is widely used for {flow phenomena}, some examples of which include: turbulent channel flow \citep{Moin1989,Ball1991}, flat-plate boundary layers \citep{Rempfer1994}, and free shear jet flows \citep{Arndt1997}. \citet{Smith2005} 
showed how to incorporate  system symmetries into the POD modes, the details of which we elaborate on in Sec.\ \ref{sec:Results}.

While the POD has seen wide use and is easy to interpret, more accurate reconstruction can be achieved with nonlinear methods -- a result we highlight in Sec.\ \ref{sec:Results}.
Some popular methods {for nonlinear dimension reduction} include kernel PCA \citep{Scholkopf1998}, diffusion maps \citep{Coifman2005a}, local linear embedding (LLE) \citep{Roweis2000}, isometric feature mapping (Isomap) \citep{Tenenbaum2000}, and t-distributed stochastic neighbor embedding (tSNE) \citep{Hinton2003}. These methods are described in more detail in \cite{Linot2021}, and an overview of {other} dimension reduction methods can be found in \cite{VanDerMaaten2009}. {One} drawback of all of these methods, however, is that they reduce the dimension, but do not immediately provide a means to move from a low-dimensional state back to the full state. {A} popular dimension reduction method without these complications is the undercomplete autoencoder \citep{Hinton2006}, which uses {a neural network (NN) to map the input data into a lower-dimensional ``latent space" and another NN to map back to the original state space.} 
We describe this structure in more detail in Sec.\ \ref{sec:Framework}. Some examples where autoencoders have been used {for flow systems} include flow around a cylinder \citep{Murata2020}, flow around a flat plate \citep{Nair2020}, Kolmogorov flow \citep{Page2021,PDJ2022}, and channel flow \citep{Milano2002}. {Although we will not pursue this approach in the present work, it may be advantageous for multiple reasons to parametrize the manifold with overlapping local representations, as done in \cite{Floryan2021}.}





After reducing the dimension, the time evolution for the dynamics can be approximated from the equations of motion or in a completely data-driven manner. The classical method is to perform a Galerkin projection wherein the equations of motion are projected onto a set of modes (e.g.\ POD modes) \citep{Holmes1998}. However, in this approach all the higher POD modes are neglected. An extension of this idea, called nonlinear Galerkin, is to assume that the time derivative of the coefficients of all of the higher modes is zero, but not the coefficients themselves \citep{Titi1990,Foias1988b,Graham1993}{; this is essentially a quasisteady state approximation for the higher modes.} This improves the accuracy, but comes at a higher computational cost than the Galerkin method, although this can be somewhat mitigated by using a postprocessing Galerkin approach \citep{GarciaArchilla1998}. \citet{Wan2018} also showed a recurrent NN (RNN) -- a NN that feeds into itself -- can be used to improve the nonlinear Galerkin approximation. {This RNN structure depends on a history of inputs, making it non-Markovian.} In addition to these linear dimension reduction approaches, an autoencoder can be used with the equations of motion in the so-called manifold Galerkin approach, which \citet{Lee2020} {developed and} applied to the viscous Burgers equation .
 
%

When the equations of motion are assumed to be unknown, and only snapshots of data are available, a number of different machine learning techniques exist to approximate the dynamics. Two of the most popular techniques are RNNs and reservoir computers. 
\citet{Vlachas2019} showed both these structures do an excellent job of {capturing} the chaotic dynamics of the Lorenz-96 equation and Kuramoto-Sivashinsky equation (KSE). For {fluid flows,} autoencoders and RNNs (specifically long-short term memory networks (LSTM)) have been used to {model} flow around a cylinders \citep{Hasegawa2020b,Eivazi2020}, pitching airfoils \citep{Eivazi2020}, bluff bodies \citep{Hasegawa2020}, and MFU plane Poiseuille flow (PPF) \citep{Nakamura2021}. Although these methods often do an excellent job of predicting chaotic dynamics, the models are not {Markovian}, so the dimension of the system also includes some history, and these models perform discrete timesteps. These two properties are undesirable, because the underlying dynamics are {Markovian} and continuous {in time}, and modeling them differently complicates {applications and interpretations} of the model. In particular, we want to use the model {for state space analyses such as determination of periodic orbits, where standard tools are available for ODEs that do not easily generalize to non-Markovian dynamic models.}


Due to these issues, we  use neural ordinary differential equations (ODE) \citep{Chen2019}. In neural ODEs, the right-hand-side (RHS) of an ODE is {represented as} a NN that is trained to reconstruct {the time evolution of the data from snapshots of training data.}
 In \cite{Linot2021} it was shown {that} this is an effective method for modeling the chaotic dynamics of the KSE{.} \citet{Rojas2021} {used} neural ODEs {to} predict the periodic dynamics of flow around a cylinder, and \citet{portwood2019turbulence} {used neural ODEs to} predict the kinetic energy and dissipation of decaying turbulence.

In this work we investigate the dynamics of MFU Couette flow. The idea behind the MFU, first introduced by  \citet{Jimenez1991}, is to reduce the simulation domain to the smallest size that sustains turbulence, thus isolating the key components of the turbulent nonlinear dynamics. Using an MFU for Couette flow at transitional Reynolds number, \citet{Hamilton1995} outlined the regeneration cycle of wall bounded turbulence called the ``self-sustaining process" (SSP), which we describe in more detail in Sec.\ \ref{sec:Results}. This system was later analyzed with coviariant Lyapunov analysis by  \citet{Inubushi2015}, who found a Lyapunov time (the inverse of the leading Lyapunov exponent) of $\sim 48$ time units.

Many low-dimensional models have been developed to recreate the dynamics of the SSP. The first investigation of this topic was by \citet{Waleffe1997a}, who developed an 8 mode model for shear flow between free-slip walls generated by a spatially sinusoidal forcing. He selected the modes based on intuition from the SSP and performed a Galerkin projection onto these modes. \citet{Moehlis2004} later added an additional mode to Waleffe's model which { enables modification of} the mean profile by the turbulence, and made some modifications to the chosen modes. In this ``MFE" model, Moehlis \emph{et al.}\ found exact coherent states, which we discuss below, that did not exist in the 8 mode model. In addition, 
\citet{Moehlis2002} also used the POD modes on a domain slightly larger than the MFU to generate POD-Galerkin models. These low-dimensional models have been used as a starting point for testing data-driven models. For example, both LSTMs \citep{Srinivasan2019} and a Koopman operator method with nonlinear forcing \citep{Eivazi2021} have been used to {attempt to} reconstruct the MFE model dynamics. \citet{BORRELLI2022} then applied these methods to PPF.

Finally, {we note that a key approach to understanding complex nonlinear dynamical phenomena, such as the SSP {of near-wall turbulence}, is through study of the underlying state space structure of fixed points and periodic orbits. In the turbulence literature these are sometimes called ``exact coherent states", or ECS \citep{Kawahara2011,Graham2021}.}
Turbulence organizes around ECS in the sense that trajectories chaotically move between different such states. The first ECS found were fixed point solutions in PCF \citep{Nagata1990}. {Following this work, \citet{Waleffe1998} was able to connect ECS of PCF and PPF to the SSP.} Later, more fixed point ECS were found in MFU PCF and visualized by \citet{Gibson2008}. 
{Unlike fixed points, which cannot capture dynamic phenomena at all, periodic orbits are able to represent key aspects of turbulent dynamics such as bursting behavior.}
\citet{Kawahara2001} found the first two periodic orbits (POs) for MFU PCF, one of which had statistics that agreed well with the SSP. Then,  \citet{Viswanath2007} found another PO and 4 new relative POs (RPOs) in this domain, and Gibson made these solutions available in \citep{Gibson2008a}, along with a handful of others. 

In {the present work,} we use autoencoders and neural ODEs {, in a method we call ``Data-driven Manifold Dynamics" (DManD) \citep{Linot2022},} to build a ROM for turbulent MFU PCF \citep{Hamilton1995}. Section \ref{sec:Framework} outlines the details of the DManD framework. We then describe the details of the Couette flow in Sec. \ref{sec:Resultsa}, the results of the dimension reduction in Sec. \ref{sec:Resultsb}, and the DManD model's reconstruction of short- and long-time statistics in Sec.~\ref{sec:Resultsc} and Sec.~\ref{sec:Resultsd}, respectively. After showing that the models accurately reproduce these statistics, we {compute} RPOs {for the model} in Sec.~\ref{sec:Resultse}{, finding several that are similar to previously known RPOs, as well as several that seem to be new.} Finally, we summarize the results in Sec.~\ref{sec:Conclusion}.

\section{Framework} \label{sec:Framework}

Here we outline our method for an ``exact" DManD modeling approach. In this sense ``exact" means all of the functions described allow for perfect reconstruction, but error is introduced in approximating these functions due to insufficient data, error in learning the functions, or error in evolving them forward in time. This differs from coarse-grained ROMs{, which} approximate the physics to generate a closed set of equations. A key component allowing DManD to be ``exact" is that we only seek to discover the evolution of trajectories after they collapse onto {an invariant} manifold {$\mathcal{M}$}. 

{In general, the training data for development of a DManD model comes in the form of snapshots  $\{u_1,u_2,...,u_M\}$, which are either the full state or measurements diffeomorphic to the full state (e.g.\ time delays \citep{Takens,Young2022}).} {Here we consider full-state data $u$ that lives in an ambient space $\mathbb{R}^d$. We generate a time series of data by evolving this state forward in time according to}
\begin{equation}
	\dfrac{du}{dt}=f(u).
\end{equation}
(In the present context, this equation represents a fully-resolved direct numerical simulation (DNS).)
With the full state, we can then define a mapping to a low-dimensional state representation
\begin{equation}
	h=\chi(u),
\end{equation} 
with $h\in\mathbb{R}^{d_h}$ {is a coordinate representation on the manifold}. Finally, we define a mapping back to the full state
\begin{equation}
	\tilde{u}=\check{\chi}(h).
\end{equation}
For data that lies on a finite-dimensional invariant manifold these functions can exactly reconstruct the state (i.e.\ $\tilde{u}=u$). However, if the dimension $d_h$ is too low, or there are errors in the approximation of these functions, then $\tilde{u}$ approximates the state. Then, with this low-dimensional state representation, we can define an evolution equation
\begin{equation}
	\dfrac{dh}{dt}=g(h).
\end{equation} \label{eq:ODENet}
The DManD model consists of the three functions $\chi$, $\check{\chi}$, and $g$. By approximating these functions, the evolution of trajectories on the manifold can be performed entirely in the manifold coordinates $h$, which requires far fewer operations than a full simulation, as $d_h\ll d$. We choose to approximate all of these functions using NNs, but {other representations could be used.}
 
First, we train $\chi$ and $\check{\chi}$ using an undercomplete autoencoder. This is a NN structure consisting of an encoder which reduces dimension ($\chi$) and a decoder that expands dimension ($\check{\chi}$). 
{As mentioned in Sec.\ \ref{sec:Intro}, a common approach to dimension reduction is to project onto a set of POD modes. POD gives the optimal linear projection in terms of reconstruction error, so we use this fact to train an encoder as the sum of POD and a {correction in the form of an} NN:
\begin{equation}
	h=\chi(u;\theta_E)=U_{d_h}^Tu+\mathcal{E}(U_r^Tu;\theta_E).\label{eq:encode}
\end{equation}
In this equation, $U_k\in\mathbb{R}^{d \times k}$ is a matrix whose {$k$ } columns are the {first $k$} POD modes {as ordered by variance}, and $\mathcal{E}$ is a NN. The first term ($U_{d_h}^Tu$) is the projection onto the leading $d_h$ POD modes, and the second term is the NN correction. The matrix $U_r$ in this term may either be a full change of basis {with no approximation} ($r=d$), or involve some dimension reduction ($d>r> d_h$).}

{For mapping back to the full state (decoding), we again sum POD with a correction
\begin{equation}
	\tilde{u}=\check{\chi}(h;\theta_E)=U_r([h,0]\MDGrevise{^T}+\mathcal{D}(h;\theta_D)).\label{eq:decode}
\end{equation}
Here, $[h,0]\MDGrevise{^T}$ is the $h$ vector zero padded to the correct size, and $\mathcal{D}$ is a NN. The first term is the POD mapping back to the full space, if there were no NNs, and the second term is a NN correction. In \citet{Linot2020} we refer to this structure as a hybrid autoencoder. In Sec.\ \ref{sec:Resultsb} we contrast this to a ``standard" autoencoder where $h=\mathcal{E}(U_r^Tu;\theta_E)$ and $\tilde{u}=U_r\mathcal{D}(h;\theta_D)$.
These hybrid autoencoder operations act as a shortcut connection on the optimal linear dimension reduction, which we \citep{Linot2020} found useful for representing the data and achieving accurate reconstruction of $u$. \citet{Yu2021} also took a similar approach with POD shortcut connections over each layer of the network.}

We {determine} the NN parameters $\theta_E$ and $\theta_D$ {by minimizing}
\begin{equation} \label{eq:LossAuto}
	L=\dfrac{1}{dK}\sum_{i=1}^K||u(t_i)-\check{\chi}(\chi(u(t_i);\theta_E);\theta_D)||^2 +\dfrac{1}{d_hK}\sum_{i=1}^K\alpha ||\mathcal{E}(U_r^Tu(t_i);\theta_E)+\mathcal{D}_{d_h}(h(t_i);\theta_D)||^2.
\end{equation}
{The first term in this loss is the mean-squared error (MSE) {of the reconstruction $\tilde{u}$}, and the second term is a penalty {that promotes accurate representation of} the leading {$d_h$} POD coefficients. In this term, $\mathcal{D}_{d_h}$ denotes the leading $d_h$ elements of the decoder output. For finite $\alpha$, the autoencoder exactly matches the first $d_h$ POD coefficients when this term vanishes.} Details of the minimization procedure are discussed in Sec. \ref{sec:Results}.

Next, we approximate $g$ using a neural ODE. A drawback of training a single dense NN for $g$ is that the resulting dynamics {may become weakly unstable, with linear growth at long times \citep{Linot2021,Linot2022}.}  To avoid this, we use a ``stabilized" neural ODE approach by adding a linear damping term onto the output of the NN, giving
\begin{equation}\label{eq:ODENet_Damp}
	g(h(t_i);\theta_g)=g_{\text{NN}}(h(t_i);\theta_g)+A h.
\end{equation}
Integrating Eq.\ \ref{eq:ODENet_Damp} forward from time $t_i$ to $t_i+\tau$ yields
\begin{equation}\label{eq:ODENet_Int}
	\tilde{h}(t_i+\tau)=h(t_i)+\int_{t_i}^{t_i+\tau}g_{\text{NN}}(h(t);\theta_g)+A h(t) dt.
\end{equation}
Depending on the situation, one may either learn $A$ from data, or fix it. Here we set it to the diagonal matrix
\begin{equation} \label{eq:Damp}
	A_{ij}=-\beta \delta_{ij}\sigma_i(h)
\end{equation} 
where $\sigma_i(h)$ is the standard deviation of the $i$th component of $h$, $\beta$ is a tunable parameter, and $\delta_{ij}$ is the Kronecker delta. This linear term attracts trajectories back to the origin, {preventing them} from moving far away from the training data.
In Sec.\ \ref{sec:Resultsd} we show that this approach drastically improves the long-time performance of these models.

We then {determine} the parameters $\theta_g$ {by minimizing} the difference between the predicted state $\tilde{h}(t_i+\tau)$ and the true state $h(t_i+\tau)${, averaged over the data:}
\begin{equation} \label{eq:LossODE}
	J=\dfrac{1}{d_hK}\sum_{i=1}^K\left(||h(t_i+\tau)-\tilde{h}(t_i+\tau)||_2^2\right).
\end{equation}
For clarity we show the specific loss we use, which sums over only a single snapshot forward in time at a fixed $\tau$. More generally, the loss can be formulated for arbitrary snapshot spacing and for multiple snapshots forward in time. To compute the gradient {of $J$ with respect to the neural network parameters $\theta_g$},
 automatic differentiation can be used to backpropagate through the {ODE} solver {that is used to compute the time integral in Eq.\ \ref{eq:ODENet_Int},} or an adjoint {problem} can be solved backwards in time \citep{Chen2019}. The adjoint method uses less memory than backpropagation, but $h$ is low-dimensional and our prediction window for training is short, so we choose to backpropagate through the solver.  

So far this approach to approximating $\chi$, $\check{\chi}$, and $g$ is general and does not directly account for the fact that the underlying equations are often {invariant} to certain symmetry operations. For example, one of the symmetries in PCF is a continuous translation symmetry {in $x$ and $z$} (i.e.\ any solution shifted to another location in the domain gives another solution). This poses an issue for training, because {in principle}, the training data must include \emph{all} these {translations} to accurately model the dynamics under {any translation}. {We discuss these and other symmetries of PCF in Sec.\ \ref{sec:Resultsa}.}

{In practice, accounting for continuous symmetries is most important along directions that sample different phases very slowly. For PCF, the mean flow is in the $x$ direction, leading to good phase sampling along this direction. However, there is no mean flow in the $z$ direction, so sampling all phases relies on the slow phase diffusion in that direction. Therefore,  we will only explicitly to account for the $z$-phase in Sec.\ \ref{sec:Results}, but in the current disucssion we  present the general framework accounting for all continuous symmetries.}

To address the issue of continuous translations, we add an additional preprocessing step to the data, using the method of slices \citep{Budanur2015,Budanur2015a} to split the state $u$ into a pattern $u_p\in \mathbb{R}^d$ and a phase $\phi\in \mathbb{R}^c$. {The number of continuous translation symmetries for which we explicitly account determines $c$.} We discuss the details of computing the pattern and the phase in Sec.\ \ref{sec:Resultsa}. Separating the pattern and phase is useful because the evolution of both the pattern and the phase only depend on the pattern. Thus, we simply replace $u$ with $u_p$ in all the above equations and then write one additional ODE for the phase
\begin{equation}
	\dfrac{d\phi}{dt}=g_\phi(h;\theta_\phi).
\end{equation}
We then fix the parameters of $g$ to evolve $h$ (from $u_p$) forward in time and use that to make a phase prediction
\begin{equation}
	\tilde{\phi}(t_i+\tau)=\phi(t_i)+\int_{t_i}^{t_i+\tau}g_\phi(h(t_i);\theta_\phi) dt.
\end{equation}
Finally, we {determine} the parameters $\theta_\phi$ to minimize the difference between the predicted phase $\tilde{\phi}(t_i+\tau)$ and the true phase $\phi(t_i+\tau)$
\begin{equation} \label{eq:LossPhi}
	J_\phi=\dfrac{1}{cK}\sum_{i=1}^K\left(||\phi(t_i+\tau)-\tilde{\phi}(t_i+\tau)||^2\right),
\end{equation}
using the method described above to compute the gradient of $J_\phi$.

\section{Results} \label{sec:Results}

\subsection{Description of Plane Couette Flow Data} \label{sec:Resultsa}
In the following sections we {apply} DManD {to} DNS of turbulent PCF in a MFU domain. Specifically, we consider the well-studied Hamilton, Kim, and Waleffe (HKW) domain \citep{Hamilton1995}. We made this selection to compare our DManD results to the analysis of the self-sustaining process in this domain, to compare our DManD results to other Galerkin-based ROMs, and to compare our DManD results to known unstable periodic solutions in this domain. 

For PCF we solve the Navier-Stokes equations
\begin{equation} \label{eq:NSE}
	\dfrac{\partial \mathbf{u}}{\partial t}+\mathbf{u}\cdot \nabla \mathbf{u}=-\nabla p+\Rey^{-1}\nabla^2 \mathbf{u}, \quad \nabla \cdot \mathbf{u}=0
\end{equation} 
for a fluid confined between two plates moving in opposite directions with the same {speed}. Eq.\ \ref{eq:NSE} is the nondimensionalized form of the equations with velocities in the streamwise $x\in[0,L_x]$, wall-normal $y\in[-1,1]$, and spanwise $z\in[0,L_z]$ directions defined as $\mathbf{u}=[u_x,u_y,u_z]$, and  pressure $p$. We solve this equation for a domain with periodic boundary conditions in $x$ and $z$ {($\mathbf{u}(0,y, z)=\mathbf{u}(L_x,y, z), \mathbf{u}(x,y, 0)=\mathbf{u}(x,y, L_z)$)} and no-slip, no-penetration boundary conditions in $y$ ($u_x(x,\pm 1, z)=\pm 1, u_y(x,\pm 1, z)=u_z(x,\pm 1, z)=0$). The complexity of the flow increases as the Reynolds number $\Rey$ increases and the domain size $L_x$ and $L_z$ increase. Here we use the HKW cell, which is at $\Rey=400$ with a domain size $[L_x,L_y,L_z]=[1.75\pi,2,1.2\pi]$ \citep{Hamilton1995}. The HKW cell is one of the {simplest flows} that sustains turbulence for extended periods of time before relaminarizing. {We chose to use this flow because it is well studied (refer to Sec.\ \ref{sec:Intro}), it isolates the SSP \citep{Hamilton1995}, and a library of ECS exist for this domain \citep{Gibson2008a}. Here we are interested in modeling the turbulent dynamics, so we will remove data upon relaminarization as detailed below.}

Eq.\ \ref{eq:NSE}, under the boundary conditions described, is invariant (and its solutions equivariant) under the discrete symmetries of point reflections about $[x,y,z]=[0,0,0]$
\begin{equation}
	\mathcal{P}\cdot[(u_x,u_y,u_z,p)(x,y,z,t)]=(-u_x,-u_y,-u_z,p)(-x,-y,-z,t)
\end{equation}
reflection about the $z=0$ plane
\begin{equation}
	\mathcal{R}\cdot[(u_x,u_y,u_z,p)(x,y,z,t)]=(u_x,u_y,-u_z,p)(x,y,-z,t)
\end{equation}
and rotation by $\pi$ about the $z$-axis
\begin{equation}
	\mathcal{RP}\cdot[(u_x,u_y,u_z,p)(x,y,z,t)]=(-u_x,-u_y,u_z,p)(-x,-y,z,t).
\end{equation}
In addition to the discrete symmetries, there are also continuous translation symmetries in $x$ and $z$
\begin{equation}
	\mathcal{T}_{\sigma_x,\sigma_z}\cdot[(u_x,u_y,u_z,p)(x,y,z,t)]=(u_x,u_y,u_z,p)(x+\sigma_x,y,z+\sigma_z,t).
\end{equation}
{We incorporate all these symmetries in the POD represesntation \citep{Smith2005}, as we discuss further in Sec.\ \ref{sec:Resultsb}.}
Then, we use the method of slices \citep{Budanur2015a} to phase align in the $z$ direction. 
{By phase aligning in $z$ we fix the location of the low-speed streak.}
Without {the alignment in $z$,} models performed poorly {because the models needed to learn how to represent every spatial shift of every snapshot.}
In what follows, we only consider phase-alignment in $z$, but we note that extending this work to phase-alignment in $x$ is straightforward. To phase-align the data, we use the first Fourier mode method-of-slices \citep{Budanur2015a}. First, we compute a phase by taking the Fourier transform of the streamwise velocity in $x$ and $z$ ($\hat{u}_x(k_x,y,k_z)=\mathcal{F}_{x,z}(u_x)$) {at a specific $y$ location ($y_1$)} to compute the phase 
\begin{equation}
	\phi=\text{atan2}(\text{imag}(\hat{u}_x(0,y_1,1)),\text{real}(\hat{u}_x(0,y_1,1))).
\end{equation}
We select $y_1$ to be one grid point off the bottom wall. Then we compute the pattern dynamics by using the Fourier shift theorem to set the phase to 0 (i.e.\ move the low-speed streak to the center of the channel)
\begin{equation}
	\mathbf{u}_p=\mathcal{F}_{x,z}^{-1}(\hat{\mathbf{u}}\exp(-ik_z\phi)).
\end{equation}

We generate turbulent PCF trajectories using the psuedo-spectral Channelflow code developed by Gibson \emph{et al.} (\citeyear{Gibson2012,Gibson2021}). In this code, the velocity and pressure fields are projected onto Fourier modes in $x$ and $z$ and Chebyshev polynomials of the first kind in $y$. These coefficients are evolved forward in time first using the multistage SMRK2 scheme \citep{Spalart1991}, then, after taking multiple timesteps, using the multistep Adams-Bashforth Backward-Differentiation 3 scheme \citep{Peyret2002}. At each timestep, a pressure boundary condition is found such that incompressibility is satisfied at the wall ($du_y/dy=0$) using the influence matrix method and tau correction developed by \citet{Kleiser1980}. 

Data was generated with $\Delta t=0.0325$ on a grid of $[N_x,N_y,N_z]=[32,35,32]$ in $x$, $y$, and $z$ for the HKW cell. Starting from random divergence-free initial conditions, we ran simulations forward for either $10,000$ xtime units or until relaminarization. Then we dropped the first $1,000$ time units as transient data and the last $1,000$ time units to avoid laminar data, and repeated with a new initial condition until we had $91,562$ time units of data stored {at intervals of one} time unit. We split this data into $80\%$ for training and $20\%$ for testing.
Finally, we preprocess the data by computing the mean $\left<{\mathbf{u}}\right>(y)$ from the training data and subtracting it from all data {$\mathbf{u}'=\mathbf{u}-\left<{\mathbf{u}}\right>$}, and then we compute the pattern $\mathbf{u}'_p$ and the phase $\phi$ as described above. The pattern $u_p$ described in Sec.\ \ref{sec:Framework} is $\mathbf{u}'_p$ flattened into a vector (i.e.\ $d=3N_xN_yN_z$). The POD and NN training use only the training data, and all comparisons use test data unless otherwise specified. 

\subsection{Dimension Reduction and Dynamic Model Construction} \label{sec:Resultsb}
\subsubsection{{Linear dimension reduction with POD: From $\mathcal{O}(10^5)$ to $\mathcal{O}(10^3)$}}

The first task in DManD for this Couette flow data is finding a low-dimensional parameterization of the manifold on which the long-time dynamics lie. We parameterize this manifold in two steps. First, we reduce the dimension down from $\mathcal{O}(10^5)$ to $502$ with the proper orthogonal decomposition (POD), and, second, we use an autoencoder to reduce the dimension down to $d_h$. The first step is simply a preprocessing step to reduce the size of the data, which reduces the number of parameters in the autoencoder.
Due to Whitney's embedding theorem \citep{Whitney1936,Whitney1944}, we know that as long as the manifold dimension is less than $251$ ($d_\mathcal{M}<251$) then this POD representation is diffeomorphic to the full state. As we show later, the manifold dimension appears to be far lower than $d_\mathcal{M}=251$, so no information of the full state should be lost with this first step. 

Proper orthogonal decomposition (POD) originates with the question of what function $\boldsymbol{\Phi}$ maximizes
\begin{equation}
	\dfrac{\left<\left|(\mathbf{u}',\boldsymbol{\Phi})\right|^2\right>}{||\boldsymbol{\Phi}||^2}.
\end{equation} 
Solutions $\boldsymbol{\Phi}^{(n)}$ to this problem satisfy the eigenvalue problem 
\begin{equation}  \label{eq:Direct}
	\sum_{j=1}^3 \int_0^{L_x}\int_{-1}^1 \int_0^{L_z}
	\left\langle u'_i(\mathbf{x}, t) u_j^{\prime*}\left(\mathbf{x}^{\prime}, t\right)\right\rangle \Phi_j^{(n)}\left(\mathbf{x}^{\prime}\right) d\mathbf{x}^{\prime}=\lambda_i \Phi_i^{(n)} (\mathbf{x})
\end{equation}
\citep{Holmes1998,Smith2005}.
Unfortunately, upon approximating these integrals, with the trapezoidal rule for example, this becomes a $d \times d$ matrix, making computation intractable. Furthermore, computing the average in Eq.\ \ref{eq:Direct}, without any modifications, results in POD modes that fail to preserve the underlying symmetries described above.
 
{In order to make this problem computationally tractable, and preserve symmetries, we apply the POD method used in \citet{Smith2005}, with the slight difference that we first subtract off the mean of state before performing the analysis. The first step in this procedure is to treat the POD modes as Fourier modes in both the $x$ and $z$ directions. Holmes \emph{et al.}\ show in \citep{Holmes1998} that for translation-invariant directions Fourier modes are the optimal POD modes. This step transforms the eigenvalue problem into
\begin{equation}  \label{eq:Direct2}
L_x L_z \sum_{j=1}^3 \int_{-1}^1\left\langle \hat{u}'_i(k_x,y^{\prime},k_z,t) \hat{u}_j^{\prime*}(k_x,y^{\prime},k_z,t)\right\rangle \varphi_{jk_xk_z}^{(n)}\left(y^{\prime}\right) d y^{\prime}=\lambda_{k_xk_z}^{(n)} \varphi_{ik_xk_z}^{(n)}(y),
\end{equation}
which reduces the $d \times d$ eigenvalue problem down to a $3 N_y \times 3 N_y$ eigenvalue problem for every wavenumber pair $(k_x,k_z)$ of Fourier coefficients. We used $5,000$ snapshots evenly sampled over the training data to compute the POD modes. Then, to account for the discrete symmetries, the data is augmented such that the mean in Eq.\ \ref{eq:Direct2} is computed by adding all the discrete symmetries of each snapshot.}

\begin{figure} 
	\centering
	\includegraphics[width=\textwidth]{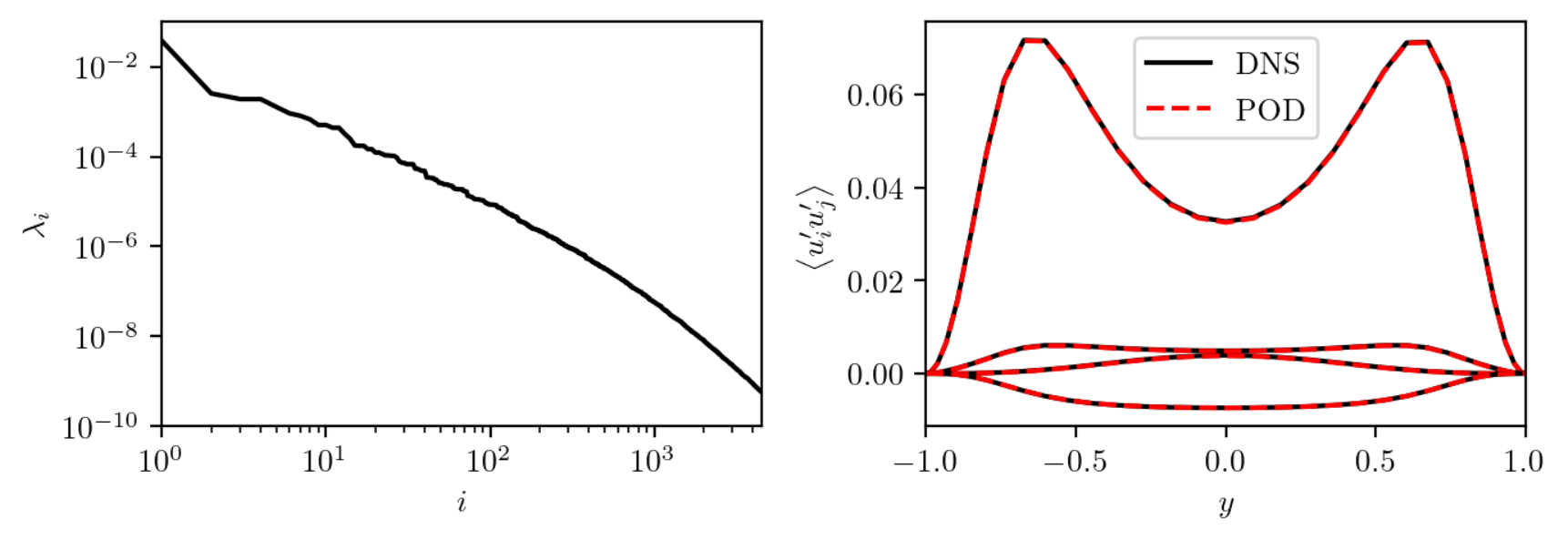}
	\captionsetup[subfigure]{labelformat=empty}
	\begin{picture}(0,0)
	\put(-185,130){\contour{white}{ \textcolor{black}{a)}}}
	\put(20,130){\contour{white}{ \textcolor{black}{b)}}}
	\end{picture} 
	\begin{subfigure}[b]{0\textwidth}\caption{}\vspace{-10mm}\label{fig:SV}\end{subfigure}
	\begin{subfigure}[b]{0\textwidth}\caption{}\vspace{-10mm}\label{fig:PODFFTa}\end{subfigure}
	
	\caption{{(a) Eigenvalues of POD modes sorted in descending order. (b) Components of the Reynolds stress for data generated by the DNS and this data projected onto 256 POD modes. In (a) the curves are, from top to bottom, $\left<u^{\prime 2}_x\right>$, $\left<u^{\prime 2}_z\right>$, $\left<u^{\prime 2}_y\right>$, and $\left<u^{\prime}_xu^{\prime}_y\right>$.}}
	\label{fig:SVPODFFT}
\end{figure}
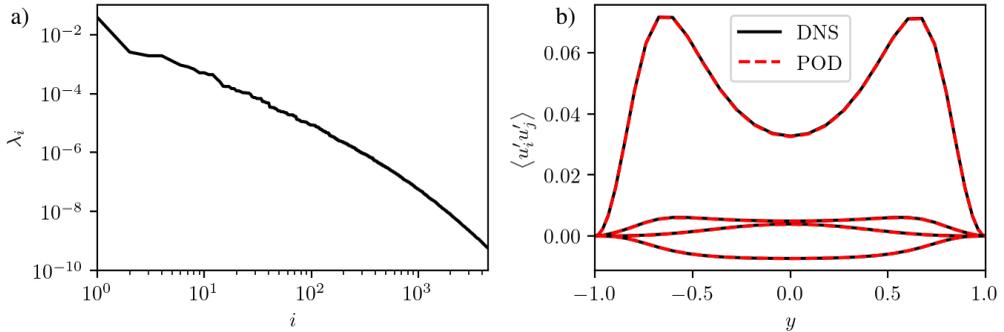

{This analysis gives us POD modes
\begin{equation}
	\boldsymbol{\Phi}_{k_xk_z}^{(n)}(\mathbf{x})=\frac{1}{\sqrt{L_x L_z}} \exp \left(2 \pi i\left(\frac{k_x x}{L_x}+\frac{k_z z}{L_z}\right)\right) \boldsymbol{\varphi}_{k_xk_z}^{(n)}(y),
\end{equation}
and eigenvalues $\lambda_{k_xk_z}^{(n)}$.
We sort the modes from largest eigenvalue to smallest eigenvalue ($\lambda_i$) and 
and project onto the leading $256$ modes, giving us {a vector of } POD coefficients $a(t)$.}  A majority of these modes are complex, so projecting onto these modes results in a $502$-dimensional system. In Fig.\ \ref{fig:SV} we plot the eigenvalues, which show a rapid drop and then a long tail that contributes little to the energy content. By dividing the {eigenvalues of the leading $256$ modes} by the total, we find these modes contain $99.8\%$ of the energy. To further illustrate that $256$ modes is sufficient to represent the state in this case, we consider the reconstruction of statistics after projecting onto the POD modes. In Fig.\ \ref{fig:PODFFTa} we show the reconstruction of four components of the Reynolds stress, $\left<u^{\prime 2}_x\right>$, $\left<u^{\prime 2}_z\right>$, $\left<u^{\prime 2}_y\right>$, and $\left<u^{\prime}_xu^{\prime}_y\right>$. The projection onto POD modes matches all of these quantities extremely well.

{Now that we have converted the data to POD coefficients and filtered out the low energy modes,}
we next train an autoencoder to {perform \emph{nonlinear} dimension reduction}. As mentioned in Sec. \ref{sec:Framework}, we phase-align the data in the spanwise direction at this step using the first Fourier mode method-of-slices. A common practice when training NNs is to normalize the data by subtracting the mean and dividing by the standard deviation of each component. We do not take this approach here because the standard deviation of the higher POD coefficients, which contribute less to the reconstruction, is much smaller than the lower POD coefficients. In order to retain the important information in the magnitudes, but put the values in a more amenable form for NN training, we instead normalize the {POD coefficients} by subtracting the mean and dividing by the maximum standard deviation. Then, we take this input and train autoencoders to minimize the loss in Eq.\ \ref{eq:LossAuto} using an Adam optimizer \citep{Kingma2015} in Keras \citep{chollet2015keras}. We train for 500 epochs with a learning rate scheduler that drops the learning rate from $10^{-3}$ to $10^{-4}$ after 400 epochs. At this point we see no improvement in the reconstruction error. For the hybrid autoencoder approach, we set $\alpha=0.01$. Table \ref{Table1} includes additional NN architecture details. 

\begin{table}
	\captionsetup{justification=raggedright}
	\caption{{Architectures of NNs. ``Shape" indicates the dimension of each layer, ``Activation" the corresponding activation functions, and ``sig" is the sigmoid activation.``Learning Rate" gives the learning rate at multiple times during training. The learning rates was dropped at even intervals.}}
	\centering
		\begin{tabular}{l*{6}{c}r}
			Function & Shape & Activation & Learning Rate \\
			\hline
			$\mathcal{E}$		& 502/1000/$d_h$ \quad & sig/linear & $[10^{-3},10^{-4}]$ \\
			$\mathcal{D}$		& $d_h$/1000/502 \quad & sig/linear & $[10^{-3},10^{-4}]$ \\
			$g_{\text{NN}}$	& $d_h$/200/200/$d_h$ \quad & sig/sig/linear & $[10^{-2},10^{-3},10^{-4}]$ \\
			$g_{\phi}$	& $d_h$/200/200/$1$ \quad & sig/sig/linear & $[10^{-2},10^{-3},10^{-4}]$ \\
		\label{Table1}
		\end{tabular}
\end{table}

\subsubsection{{Nonlinear dimension reduction with autoencoders: From $\mathcal{O}(10^3)$ to $\mathcal{O}(10^1)$}}

\begin{figure} 
	\centering
	\includegraphics[width=.5\textwidth]{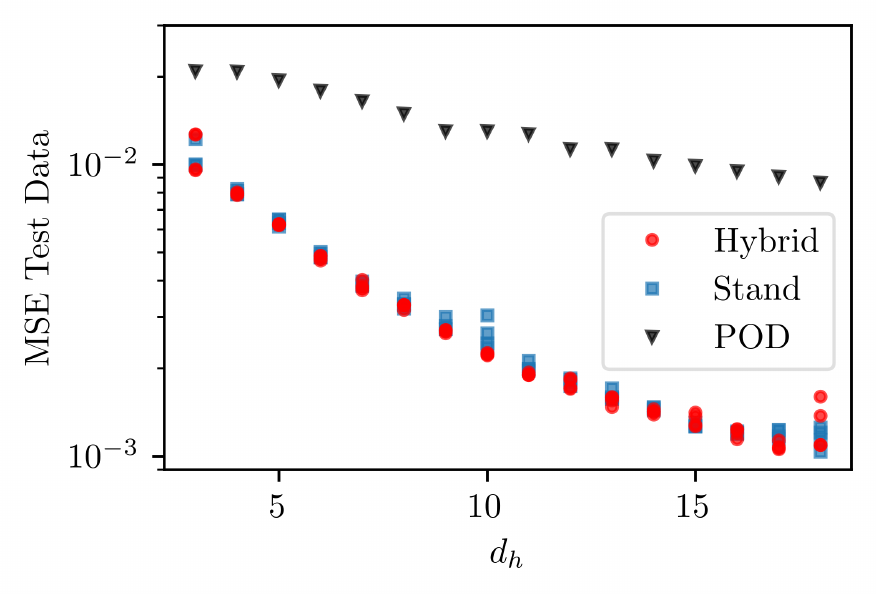}
	\caption{Mean squared error on test data for POD, standard autoencoders, and hybrid autoencoders at various dimensions $d_h$. At each dimension there are four standard and four hybrid autoencoders.}
	\label{fig:Comparison}
\end{figure}

{With the above ``preprocessing" step completed, we now turn to the reduction of dimension with the nonlinear approach enabled by the autoencoder structure.}
{We consider three approaches to reducing the dimension of $a$: (1)  Training a hybrid autoencoder, (2) Training a standard autoencoder, (3)  linear projection onto a small set of POD modes. We describe the first two approaches in Sec.\ \ref{sec:Framework}, noting that the POD projection onto $256$ (complex) modes can be written as $a=U_r^Tu$. 
{The third approach just corrsponds to setting $\mathcal{E}$ and $\mathcal{D}$ to zero in Eqs.~\ref{eq:encode} and \ref{eq:decode}.}
In Fig.\ \ref{fig:Comparison} we show the MSE of reconstructing $a$ with these three approaches over a range of different dimensions $d_h$. We use NNs with the same architectures for both the standard and the hybrid autoencoder approaches. Due to the variability introduced into autoencoder training by randomly initialized weights and stochasticity in the optimization, we show the error for four separately trained autoencoders, at each $d_h$.
 We see that the autoencoders perform an order magnitude better than POD in the range of dimension considered here. Both the standard and hybrid autoencoder approaches perform the same, so we select the hybrid approach because it 
 {can be viewed as a nonlinear correction to the POD projection}. Next we use the low-dimensional representations from these autoencoders to train stabilized neural ODEs.}
 
\subsubsection{Neural ODE Training} \label{sec:Resultsc1}

After training four autoencoders at each dimension $d_h$, we {chose a set of damping parameters, $\beta$, and for each, then trained four stabilized neural ODEs for all four autoencoders at every dimension $d_h$. This results in $16$ models at every $d_h$ and $\beta$. The final $\beta$ value of $0.1$ was selected so that long-time trajectories neither blew up nor decayed too strongly.} 
{Before training the ODEs, we preprocess each autoencoder's latent space data set $h$ by subtracting the mean.} {
It is important to center the data because the linear damping (Eq.\ \ref{eq:Damp}) pushes trajectories towards the origin.} We train the stabilized neural ODEs to predict the evolution of the centered data by using an Adam optimizer in Pytorch \citep{Paszke2019,Chen2019} to minimize the loss in Eq.\ \ref{eq:LossODE}. 
 We train using a learning rate scheduler that drops at three even intervals during training and we train until the learning curve stops improving. Table \ref{Table1} shows the details of this NN.  
 Unless otherwise stated, we show results for {the} one model out of those sixteen at each dimension with the lowest relative error averaged over all the statistics we consider. 

\subsection{Short-time tracking} \label{sec:Resultsc}

\begin{figure} 
	\centering
	\includegraphics[width=\textwidth]{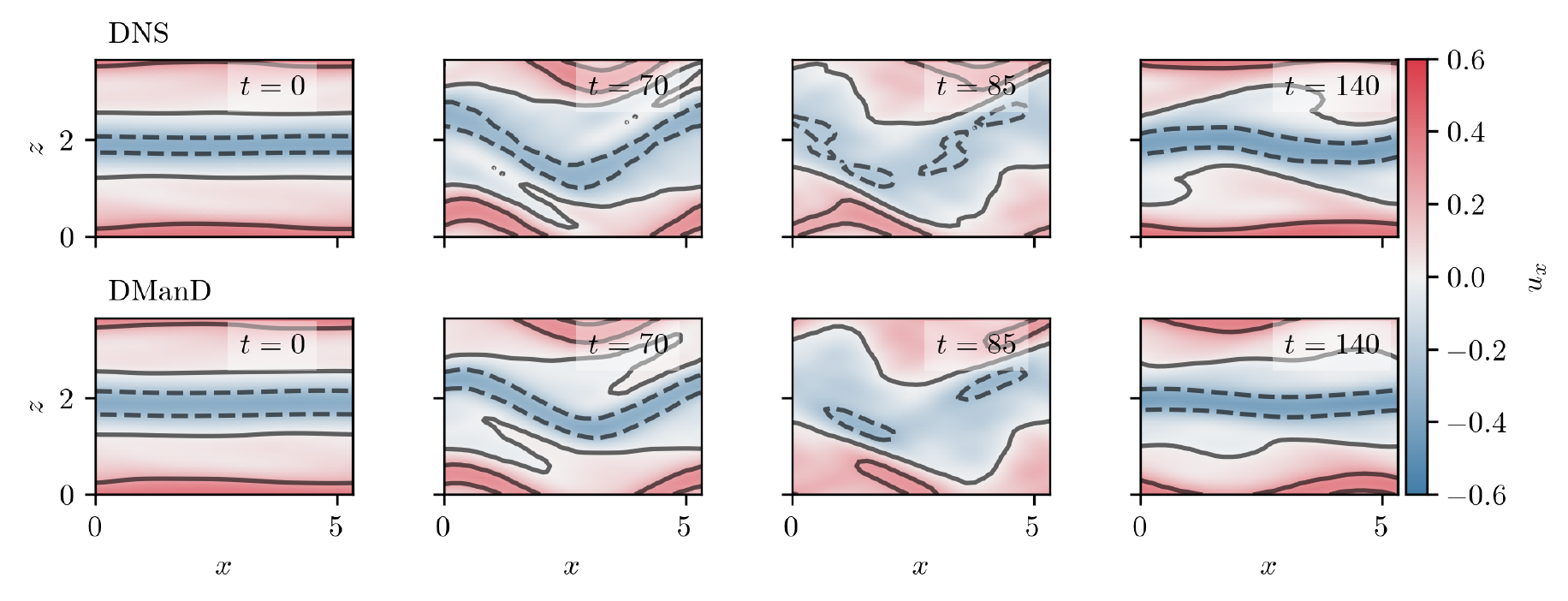}
	\caption{Snapshots of the streamwise velocity at $y=0$ from the DNS and from the DManD model at $d_h=18$.}
	\label{fig:SSPsnaps}
\end{figure}

In the following two sections we evaluate the performance of the DManD models at reconstructing short-time trajectories and long-time statistics. Figure \ref{fig:SSPsnaps} shows snapshots of the streamwise velocity at the center plane of the channel, $y=0$, for the DNS and DManD at $d_h=18$.
{We choose to show results for $d_h=18$ because the autoencoder error begins to level off around this dimension, and, as we will show, the error in statistics levels off before this dimension. The value $d_h=18$ is not necessarily the minimal dimension required to model this system.}
{In Fig.\ \ref{fig:SSPsnaps}, both the DNS and the DmanD model show key characteristics of the SSP: (1) low-speed streaks become wavy, (2) the wavy low-speed streaks break down generating rolls, (3) the rolls lift fluid from the walls, regenerating streaks.}

\begin{figure} 
	\centering
	\includegraphics[width=.5\textwidth]{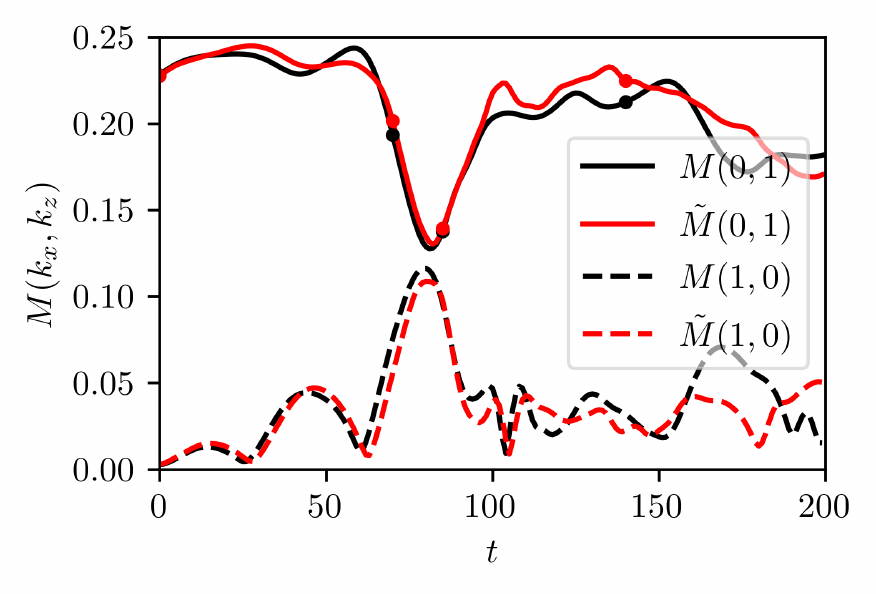}
	\caption{Modal RMS velocity from the DNS ($M$) and from the DManD model at $d_h=18$ ($\tilde{M}$). The markers correspond to the times in Fig. \ref{fig:SSPsnaps}.}
	\label{fig:SSPCycle}
\end{figure}

{Not only does DManD capture the qualitative behavior of the SSP, but Fig.\ \ref{fig:SSPsnaps} also shows good quantitative agreement as well.}
To further illustrate this, in Fig.\ \ref{fig:SSPCycle} we show the modal root-mean squared (RMS) velocity
\begin{equation}
	M(k_x,k_z)=\left(\int_{-1}^1 (\hat{u}^2_x(k_x,y,k_z)+\hat{u}^2_y(k_x,y,k_z)+\hat{u}^2_z(k_x,y,k_z))dy\right)^{1/2},
\end{equation}
which \citet{Hamilton1995} used to identify the different parts of the SSP. Specifically, we consider the $M(0,1)$ mode, which corresponds to the low speed streak and the $M(1,0)$ mode which corresponds to the $x$-dependence that appears when the streak becomes wavy and breaks up. In this example, the two {curves} match well over a cycle of the SSP and {only} start to move away after $\sim150$ time units{, which is about three Lyapunov times.}

\begin{figure} 
	\centering
	\includegraphics[width=.5\textwidth]{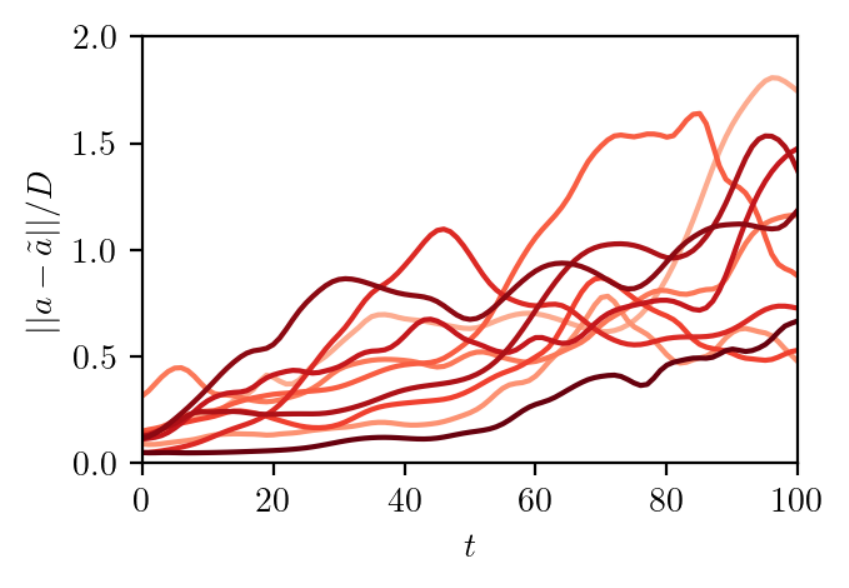}
	\caption{Normalized tracking error for 10 random initial conditions {(different shades)} using  DManD with $d_h=18$.}
	\label{fig:ExTrack}
\end{figure}

While the previous result shows a single example, we also consider ensembles of initial conditions. Figure \ref{fig:ExTrack} shows the tracking error 
 $||a(t_i+t)-\tilde{a}(t_i+t)||$ of 10 trajectories, starting at $t_i$, for a model at $d_h=18$. Here we normalize the tracking error by the error between solutions at random times $t_i$ and $t_j$  $D=\left<||a(t_i)-a(t)||\right>$. In this case the darkest line corresponds to the flow field in Figs.\ \ref{fig:SSPsnaps} and \ref{fig:SSPCycle}. 
 When considering {the} other initial conditions in Fig.\ \ref{fig:ExTrack}, there tends to be a relatively slow rise in the error over $\sim$50 time units and then a more rapid increase after this point. To better understand how this tracking varies with the dimension of the model we next consider the ensemble-averaged tracking {error}.
 
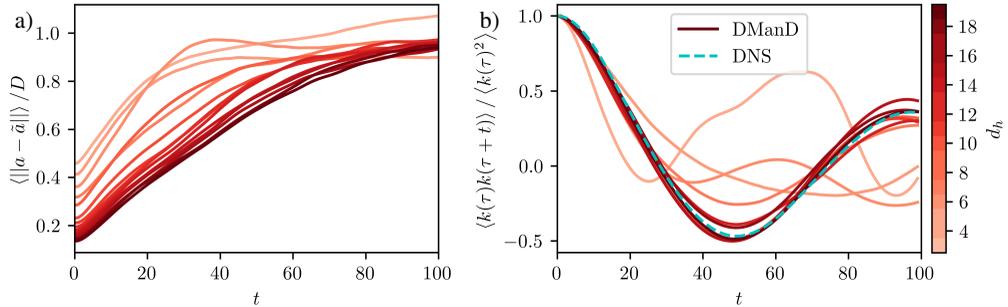
\begin{figure} 
	\centering
	\includegraphics[width=\textwidth]{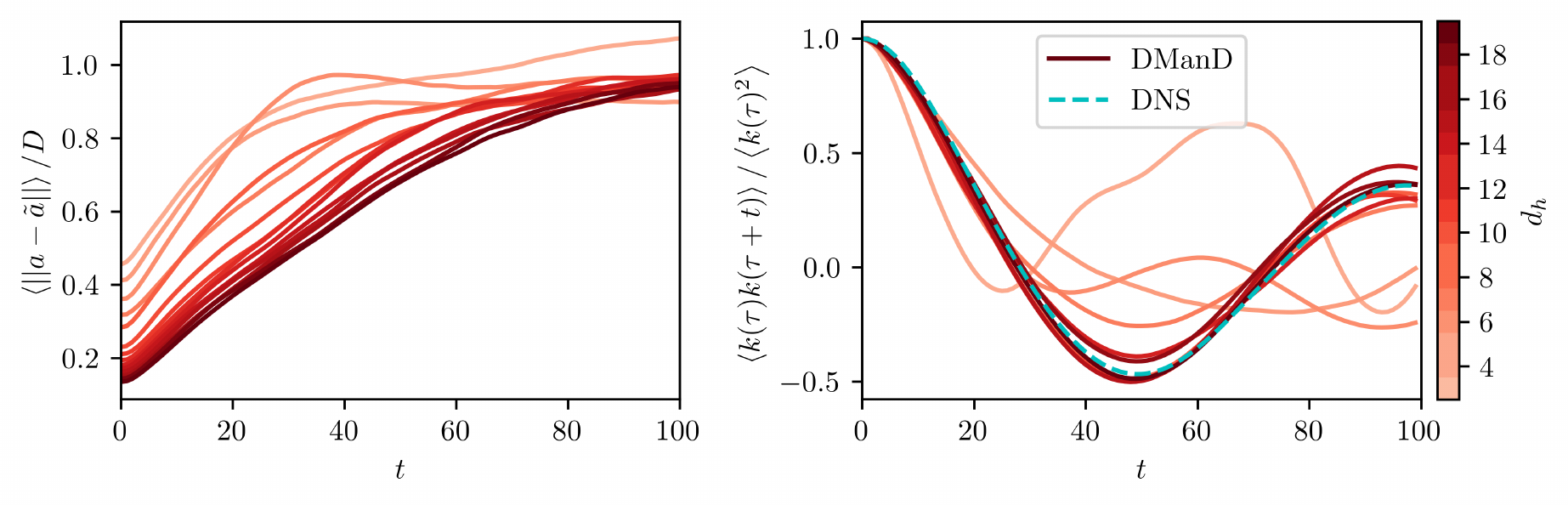}
	\captionsetup[subfigure]{labelformat=empty}
	\begin{picture}(0,0)
	\put(-185,120){\contour{white}{ \textcolor{black}{a)}}}
	\put(-10,120){\contour{white}{ \textcolor{black}{b)}}}
	\end{picture} 
	\begin{subfigure}[b]{0\textwidth}\caption{}\vspace{-10mm}\label{fig:Trackinga}\end{subfigure}
	\begin{subfigure}[b]{0\textwidth}\caption{}\vspace{-10mm}\label{fig:Trackingb}\end{subfigure}
	
	\caption{(a) ensemble averaged short-time tracking and (b) temporal autocorrelation of the kinetic energy for DmanD models of increasing dimension. In (b) odd numbers above $d_h=5$ are omitted for clarity.}
	\label{fig:Tracking}
\end{figure}

In Fig.\ \ref{fig:Trackinga} we show the normalized ensemble-averaged tracking error for model {dimensions between $d_h=3$ and $18$}. For $d_h=3-5$ there is a rapid rise in the error until $\sim$40 time units after which the error {levels off}. This behavior often happens due to trajectories quickly diverging and landing on stable fixed points or periodic orbits that do not exist in the true system. For $d_h=6-10$ there is an intermediate behavior where lines diverge more quickly than the higher-dimensional models, but tend to approach the same tracking error at $\sim$100 time units. Then, for the remaining models $d_h=11-18$, there is a smooth improvement in the tracking error over this time interval. As the dimension increases in this range the trends stay the same, but the error continues to {decrease}, which is partially due to improvement in the autoencoder performance.

The instantaneous kinetic energy of the flow is 
\begin{equation} \label{eq:KE}
	 E(t)=\frac{1}{2L_x L_z} \int_0^{L_z} \int_{-1}^1 \int_0^{L_x} \frac{1}{2}\mathbf{u}\cdot\mathbf{u} dx dy dz,
	 \end{equation}
	 and we denote its fluctuating part as $k(t)=E(t)-\left<E\right>$.
 In Fig.\ \ref{fig:Trackingb} we show the temporal autocorrelation of $k$. Again, for $d_h=3-5$ we see clear disagreement between the true autocorrelation and the prediction. Above $d_h>5$ all of the models match the temporal autocorrelation well, without a clear trend in the error as dimension changes. All these models match well for $\sim40$ time units, with $d_h=18$ (the darkest line) matching the data extremely well for two Lyapunov times. 

Finally, before investigating the long-time predictive capabilities of the model, we show the tracking of phase dynamics for $d_h=18$. As mentioned in Sec. \ref{sec:Framework}, we decouple the phase and pattern dynamics such that the time evolution of the phase only depends upon the pattern dynamics. Here we take the $d_h=18$ model and used it to train an ODE for the phase dynamics. For training we repeat the process used for training $g_{NN}$ to train $g_\phi$ with the loss in Eq.\ \ref{eq:LossPhi}. Table \ref{Table1} contains details on the NN architecture.

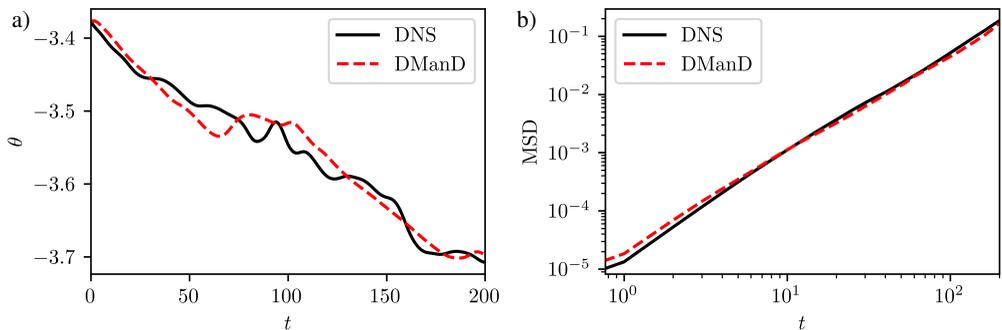
\begin{figure} 
	\centering
	\includegraphics[width=\textwidth]{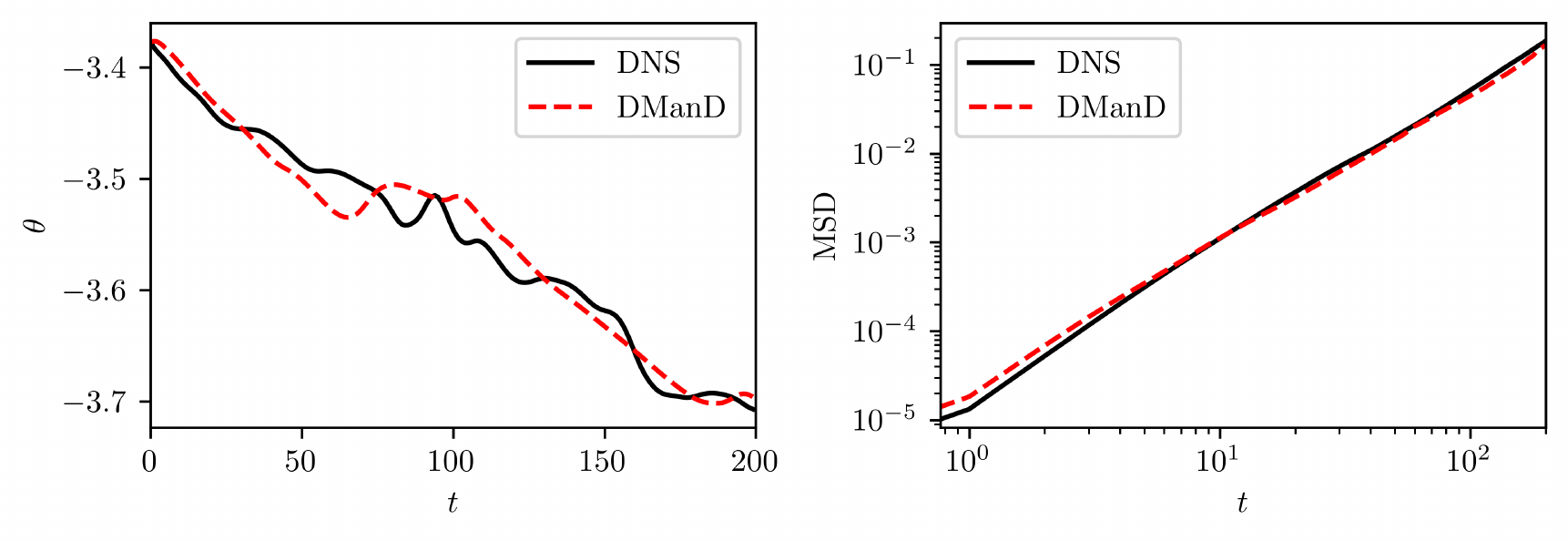}
	\captionsetup[subfigure]{labelformat=empty}
	\begin{picture}(0,0)
	\put(-185,130){\contour{white}{ \textcolor{black}{a)}}}
	\put(5,130){\contour{white}{ \textcolor{black}{b)}}}
	\end{picture} 
	\begin{subfigure}[b]{0\textwidth}\caption{}\vspace{-10mm}\label{fig:Phasea}\end{subfigure}
	\begin{subfigure}[b]{0\textwidth}\caption{}\vspace{-10mm}\label{fig:Phaseb}\end{subfigure}

	\caption{(a) an example of the phase evolution and (b) the MSD of the phase evolution for the DNS and the DManD model at $d_h=18$.}
	\label{fig:Phase}
\end{figure}

In Fig.\ \ref{fig:Phasea} we show an example of the model phase evolution over 200 time units. In this example, the model follows the same downward drift in phase despite not matching exactly. Then, to show the statistical agreement between the DNS and the model, we show the mean squared {phase} displacement $\text{MSD}=\left<(\theta(t_i)-\theta(t_i+t))^2\right>$ for both the DNS and the model in Fig.\ \ref{fig:Phaseb}, as was done for Kolmogorov flow by \cite{PDJ2022}. The curves are in good agreement.
All of the remaining long-time statistics {we report} are phase invariant, so the remaining results use only models for the pattern dynamics.

\subsection{Long-time statistics} \label{sec:Resultsd}

Next we investigate the ability of the DManD model to capture the long-time dynamics of PCF. An obvious prerequisite for models to capture long-time dynamics is the long-time stability of the models. As mentioned in Sec.\ \ref{sec:Framework}, the long-time trajectories of standard neural ODEs often become unstable, which led us to use stabilized neural ODEs with an {explicit} damping term. We quantify this observation by counting, of the $16$ models trained at each dimension $d_h$, how many become unstable with and without the presence of an explicit damping term. From our training data we know where $h$ should lie, so if it falls far outside this range after some time we can classify the model as unstable. In particular, we classify models as unstable if the norm of the final state is two times that of the maximum in our data ($||\tilde{h}(T)||>2\max_t ||h(t)||$), after $T=10^4$ time units. In all of the unstable cases $||\tilde{h}(t)||$ follows the data over some short time range before eventually {growing} indefinitely. 

\begin{figure} 
	\centering
	\includegraphics[width=.5\textwidth]{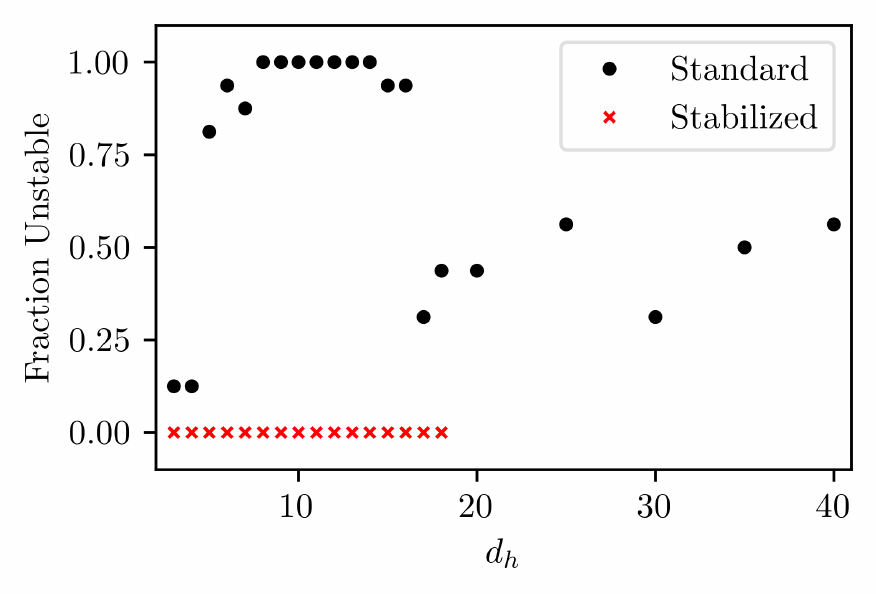}
	\caption{Fraction of unstable DManD models with standard neural ODEs and with stabilized neural ODEs at various dimensions.}
	\label{fig:Stab}
\end{figure}

In Fig.\ \ref{fig:Stab} we show the number of unstable models with and without damping. With damping, all of the models are stable, whereas without damping almost all models become unstable for $d_h=5-16$, and around half become unstable in the other cases. Additionally, with longer runs or with different initial conditions, many of the models without damping labelled as stable here also eventually become unstable. 
This lack of stability happens when inaccuracies in the neural ODE model pushes trajectories off the attractor. Once off the attractor, the model is presented with states unlike the training data leading to further growth in this error. In \citet{Linot2021,Linot2022} we show more results highlighting this behavior.
So, although some standard neural ODE models do provide reasonable statistics, using these models presents challenges due to this lack of robustness. As such, all other results we show use stabilized neural ODEs.

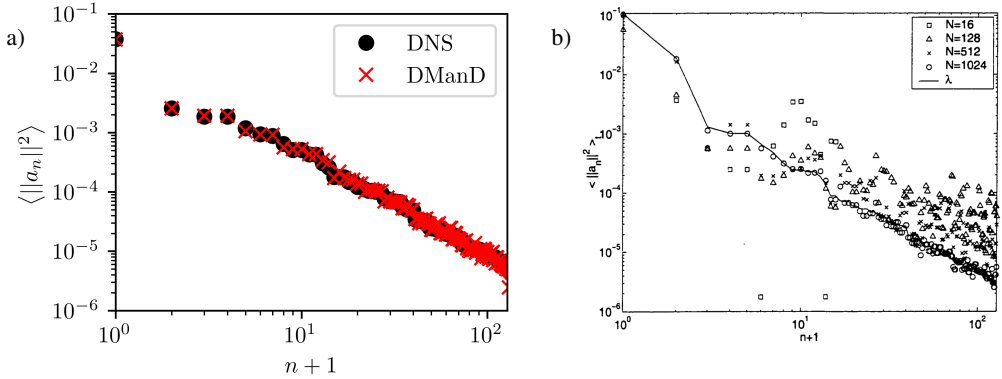
\begin{figure} 
	\centering
	\includegraphics[trim=0 0 0 0,width=\textwidth]{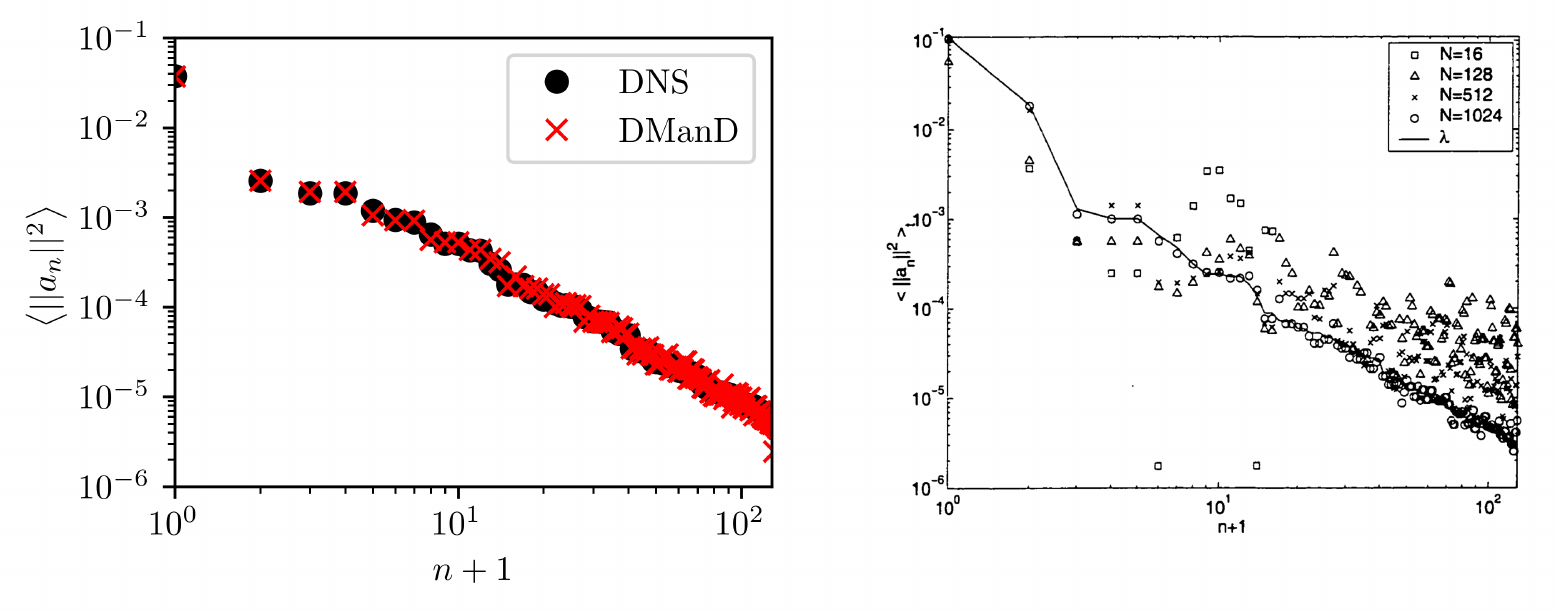}
	\captionsetup[subfigure]{labelformat=empty}
	\begin{picture}(0,0)
	\put(-190,140){\contour{white}{ \textcolor{black}{a)}}}
	\put(15,140){\contour{white}{ \textcolor{black}{b)}}}
	\end{picture}
	\begin{subfigure}[b]{0\textwidth}\caption{}\vspace{-10mm}\label{fig:GibCompa}\end{subfigure}
	\begin{subfigure}[b]{0\textwidth}\caption{}\vspace{-10mm}\label{fig:GibCompb}\end{subfigure}
	\caption{Comparison of  $\left<||a_n||^2\right>$ (mean-squared POD coefficient amplitudes) from the DNS to (a) $\left<||a_n||^2\right>$ from the DManD model at $d_h=18$ and (b) $\left<||a_n||^2\right>$ from POD-Galerkin with $N$ POD modes (reproduced with permission from \citet{Gibson2002}). In (b), the quantity $\lambda$ is equivalent to $\left<||a_n||^2\right>$ from the DNS.}
	\label{fig:GibComp}
\end{figure}

While Fig.\ \ref{fig:Stab} indicates that stabilized neural ODEs predict $\tilde{h}$ in a similar range to that of the data, it does not quantify the accuracy of these predictions. In fact, with few dimensions many of these models do not remain chaotic, landing on fixed points or periodic orbits. The first metric we use to quantify the long-time performance of the DManD method is the mean-squared POD coefficient amplitudes ($\left<||a_n||^2\right>$).
We consider this quantity because Gibson reports it for POD-Galerkin in \citet{Gibson2002} at various levels of truncation. In Fig.\ \ref{fig:GibComp} we show how well {the} DManD model{, with $d_h=18$,} captures this quantity{, in comparison to} the POD-Galerkin model in \citet{Gibson2002}. The two {data} sets slightly differ because we subtract the mean before applying POD and Gibson {did not}. The DManD method, with only 18 degrees of freedom, matches the mean-squared amplitude{s to high accuracy, far better} than all of the POD-Galerkin models. It is not until POD-Galerkin keeps 1024 modes that the results become comparable, which corresponds to $\sim 2000$ degrees of freedom because most coefficients are complex. Additionally, our method requires only data, whereas the POD Galerkin approach requires both data for computing the POD and knowledge of the equations of motion for projecting the equations onto these modes.

\begin{figure} 
	\centering
	\includegraphics[trim=0 0 0 0,width=\textwidth]{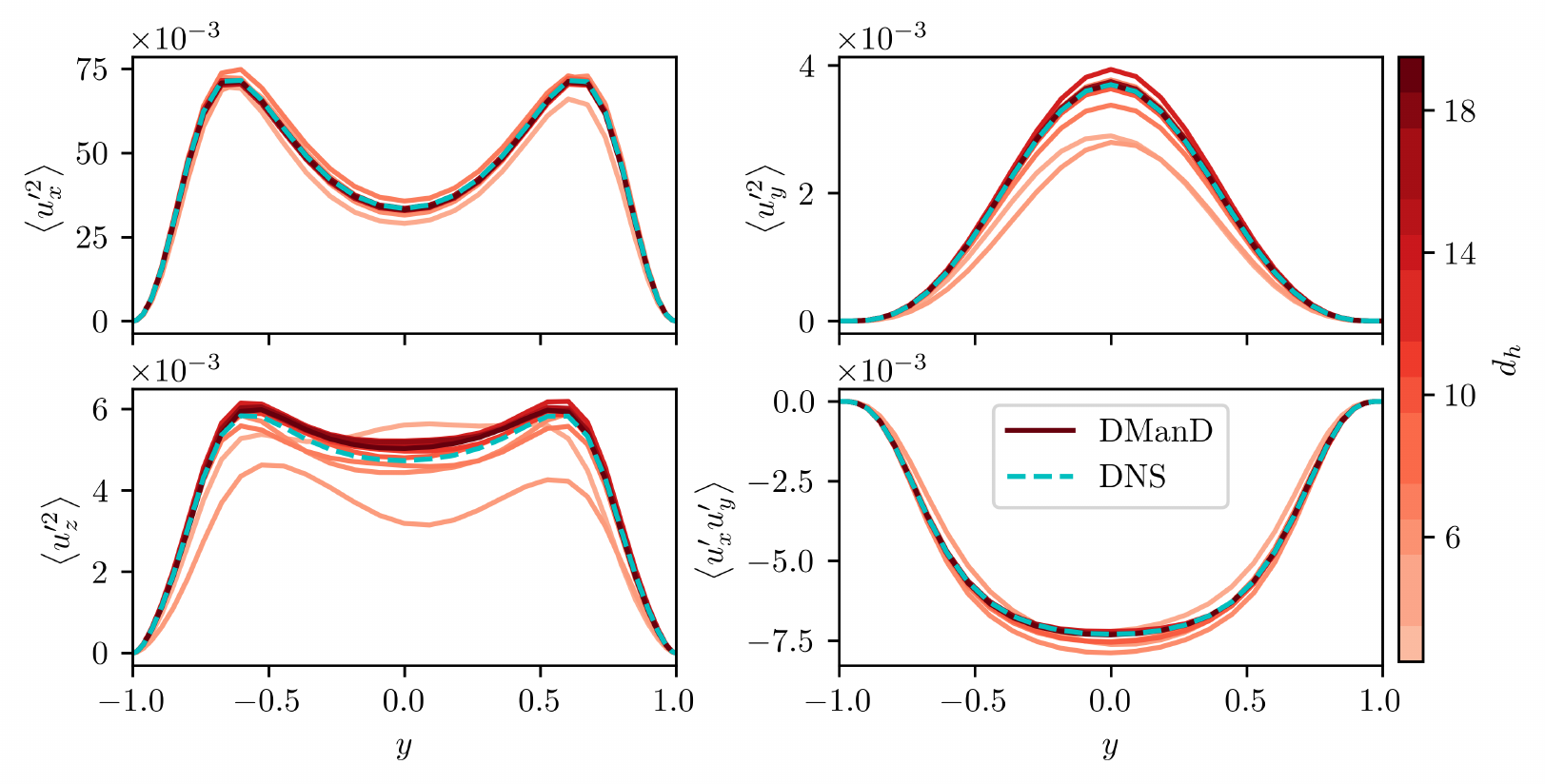}
	\caption{Components of the Reynolds stress with increasing dimension for DManD models at various dimensions. Odd numbers above $d_h=5$ are omitted for clarity.}
	\label{fig:ReStress}
\end{figure}

We now investigate how the Reynolds stress and the power input vs.\ dissipation vary with dimension. Figure \ref{fig:ReStress} shows four {components} of the Reynolds stress at various dimensions. For $\left<u_x^{\prime 2}\right>$ and $\left<u'_xu'_y\right>$, nearly all the models match the data, with relatively small deviations only appearing {for} $d_h\sim 3-6$. For $\left<u_y^{\prime 2}\right>$ and $\left<u_z^{\prime 2}\right>$, this deviation becomes more obvious, and the lines do not converge until around $d_h > 10$, with all models {above} this dimension exhibiting a minor overprediction in $\left<u_z^{\prime 2}\right>$.

To evaluate how accurate the models are at reconstructing the energy balance, we look at {joint} PDFs of {power input and dissipation.}
The power input is the amount of energy required to move the walls:
\begin{equation}
	I=\dfrac{1}{2L_xL_z}\int_0^{L_x}\int_0^{L_z} \left.\dfrac{\partial u_x}{\partial y}\right|_{y=-1}+\left.\dfrac{\partial u_x}{\partial y}\right|_{y=1}dxdz,
\end{equation}
and the dissipation is the energy lost to heat due to viscosity:
\begin{equation}
	D=\dfrac{1}{2L_xL_z}\int_0^{L_x}\int_{-1}^1\int_0^{L_z} \left| \nabla \times \mathbf{u}\right|^2 dxdydz.
\end{equation} 
These two terms define the rate of change of energy in the system $\dot{E}=I-D$, which must average to  zero over long times. Checking this statistic is important to show the DManD models correctly balance the energy. 

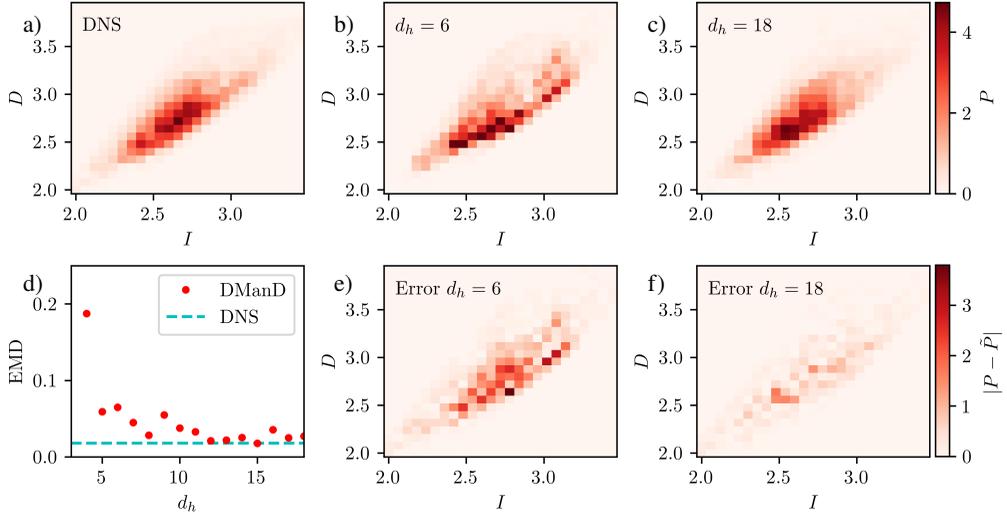
\begin{figure} 
	\centering
	\includegraphics[width=\textwidth]{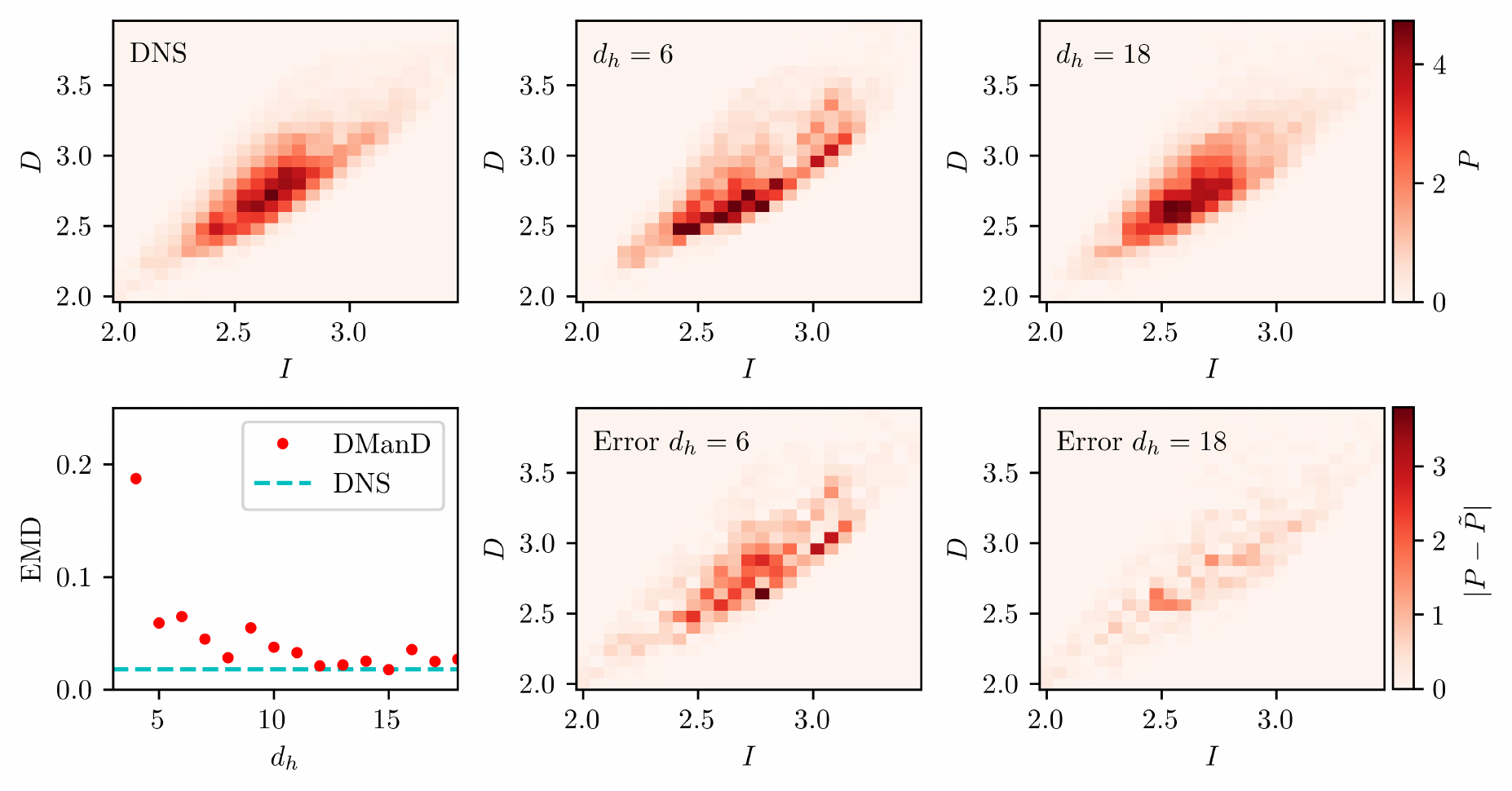}
	\captionsetup[subfigure]{labelformat=empty}
	\begin{picture}(0,0)
	\put(-177,195){\contour{white}{ \textcolor{black}{a)}}}
	\put(-60,195){\contour{white}{ \textcolor{black}{b)}}}
	\put(58,195){\contour{white}{ \textcolor{black}{c)}}}
	\put(-177,98){\contour{white}{ \textcolor{black}{d)}}}
	\put(-60,98){\contour{white}{ \textcolor{black}{e)}}}
	\put(58,98){\contour{white}{ \textcolor{black}{f)}}}
	\end{picture}
	\begin{subfigure}[b]{0\textwidth}\caption{}\vspace{-10mm}\label{fig:PDFa}\end{subfigure}
	\begin{subfigure}[b]{0\textwidth}\caption{}\vspace{-10mm}\label{fig:PDFb}\end{subfigure}
	\begin{subfigure}[b]{0\textwidth}\caption{}\vspace{-10mm}\label{fig:PDFc}\end{subfigure}
	\begin{subfigure}[b]{0\textwidth}\caption{}\vspace{-10mm}\label{fig:PDFd}\end{subfigure}
	\begin{subfigure}[b]{0\textwidth}\caption{}\vspace{-10mm}\label{fig:PDFe}\end{subfigure}
	\begin{subfigure}[b]{0\textwidth}\caption{}\vspace{-10mm}\label{fig:PDFf}\end{subfigure}
	\caption{(a)-(c): examples of joint PDFs for the true system, the DManD model at $d_h=6$, and the DManD model at $d_h=18$. (d): earth movers distance between the PDF from the DNS and the PDFs predict by the DManD model at various dimensions. ``DNS" is the error between two PDFs generated from DNS trajectories of the same length with different initial conditions. (e) and (f): the error associated with the DManD model PDFs at $d_h=6$ and $d_h=18$.}
	\label{fig:PDF}
\end{figure}

Figures \ref{fig:PDFa}-\ref{fig:PDFc} show the PDF from the DNS, the PDF for $d_h=6$ and the PDF for $d_h=18$, generated from a single trajectory evolved for $5000$ time units, and Figs.\ \ref{fig:PDFe} and \ref{fig:PDFf} show the {the absolute difference between the true and model PDFs}. With $d_h=6$ the model overestimates the number of low dissipation states, while $d_h=18$ matches the density well. In Fig.\ \ref{fig:PDFd} we compare the joint PDFs at all dimension  with the true PDF {using} the earth movers distance (EMD) \citep{Rubner1998}. The EMD determines the distance between two PDFs as a solution to the \emph{transportation problem} by treating the true PDF as ``supplies" and the model PDF as ``demands" and finding the flow which minimizes the work required to move one to the other. 
{We compute the distance between PDFs using the EMD because it is a \emph{cross-bin} distance, meaning the distance accounts for the density in neighboring bins. This is in contrast to \emph{bin-to-bin} distances, like the KL divergence, which only uses the error at a given bin. Bin-to-bin distances can vary significantly with small shifts in one PDF (misalignment) and when changing the number of bins used to generate the PDF \citep{Ling2007}. We choose the EMD because it does not suffer from these issues.}
In Fig.\ \ref{fig:PDFd} we see a steep drop in the EMD at $d_h=5$ and after $d_h>10$ the joint PDFs are in excellent agreement with the DNS. {The dashed line corresponds to the EMD between two different trajectories from the DNS.}

\subsection{Finding ECS in the model} \label{sec:Resultse}

Now that we know that {the} DManD model quantitatively captures many of the key characteristics of MFU PCF, we now want to explore using the model to discover ECS. In particular, we first investigate the whether known periodic orbits of the DNS exist in the DManD model, and then we use the DManD model to search for new periodic orbits.
Here we note that because our model predicts phase-aligned dynamics, the periodic orbits of the DManD model are either periodic or relative periodic orbits{, depending on the phase evolution, which we have not tracked}. In the following we omit all $\tilde{\cdot}$, so all functions should be assumed to come from a DManD model. 

{Here we outline the approach we take to find periodic orbits, which follows \citet{ChaosBook}. When searching for periodic orbits we seek an initial condition to a trajectory that repeats after some time period. This is equivalent to finding the zeros of 
\begin{equation} \label{eq:zeros}
	H(h,T)=G_T(h)-h,
\end{equation} 
where $G_T(h)$ is the flow map forward $T$ time units from $h$: i.e.~$G_T(h(t))=h(t+T)$. We compute $G_T(h)$ from Eq.\ \ref{eq:ODENet_Int}. Finding zeros to Eq.\ \ref{eq:zeros} requires that we find both a point $h^*$ on the periodic orbit and a period $T^*$ such that $H(h^*,T^*)=0$. One way to find $h^*$ and $T^*$ is by using the Newton-Raphson method.}

By performing a Taylor series expansion of $H$ we find near the fixed point $h^*, T^*$ of $H$ that
\begin{equation}
	\begin{aligned}
		H(h^*, T^*)-H(h, T) &\approx D_{h} H(h, T) \Delta h+D_T H(h, T) \Delta T \\
		-H(h, T) &\approx D_{h} H(h, T) \Delta h+g\left(G_T(h)\right) \Delta T,
	\end{aligned}
	\label{eq:PO1}
\end{equation} 
where $D_{h}$ is the Jacobian of $H$ with respect to $h$, $D_T$ is the Jacobian of $H$ with respect to the period $T$, $\Delta h=h^*-h$ and $\Delta T=T^*-T$. {To have a complete set of equations for $\Delta h$ and $\Delta T$, we supplement Eq.~\ref{eq:PO1} with the constraint that the updates $\Delta h$ are orthogonal to the vector field at $h$: i.e., 
\begin{equation}
	g(h)^T \Delta h=0. 
\end{equation}}
 With this constraint, at  Newton step $(i)$, the system of equations becomes 
\begin{equation}
	\begin{aligned}
	&\left[\begin{array}{cc}
	D_{h^{(i)}} H(h^{(i)}, T^{(i)}) & g(G_{T^{(i)}}(h^{(i)})) \\
	g(h^{(i)})^T& 0
	\end{array}\right]
	\left[\begin{array}{c}
	\Delta h^{(i)} \\
	\Delta T^{(i)}
	\end{array}
	\right]
	 =-\left[\begin{array}{c}
	H(h^{(i)},T^{(i)}) \\
	0
	\end{array}\right],
	\end{aligned}
	\label{eq:PO2}
\end{equation}
 {which, in the standard Newton-Raphson method, is used to update the guesses $h^{(i+1)}=h^{(i)}+\Delta h^{(i)}$ and $T^{(i+1)}=T^{(i)}+\Delta T^{(i)}$.}\

{Typically, a Newton-Krylov method is used to avoid explicitly constructing the Jacobian \citep{Viswanath2007}.}
However, with DManD, computing the Jacobian is simple, fast, and requires little memory {because the state representation is dramatically smaller in the DManD model than in the DNS}. We compute the Jacobian $D_{h} H(h, T)$ directly by using the same automatic differentiation tools used for training the neural ODE.
{Furthermore, if we had chosen to represent the dynamics in discrete, rather than continuous time, computation of general periodic orbits would not be possible, as the period $T$ can take on arbitrary values and a discrete-time representation would limit $T$ to multiples of the time step.}
When finding periodic orbits of the DManD model we used the Scipy ``hybr" method, which uses a modification of the Powell hybrid method \citep{Scipy}, and for finding periodic orbits of the DNS we used {the} Newton GMRES-Hookstep method built into Channelflow \citep{Gibson2021}. {In the following trials we only consider DManD models with $d_h=18$.}

\begin{figure} 
	\centering
	\includegraphics[width=\textwidth]{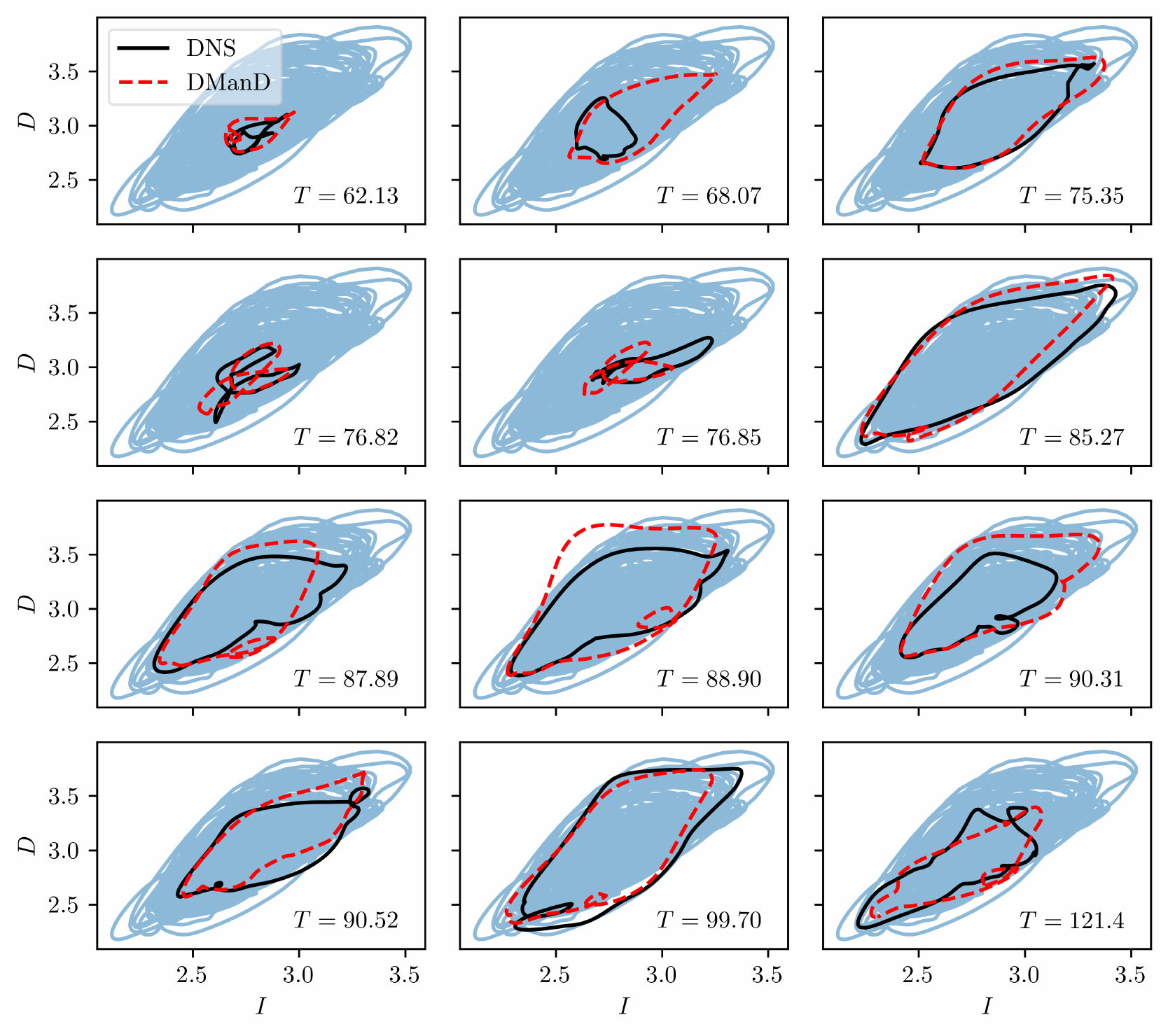}
	\caption{Power input vs.\ dissipation of known periodic orbits (period reported in bottom right) from the DNS and periodic orbits found in the DManD model at $d_h=18$. The blue line is a long trajectory of the DNS for comparison.}
	\label{fig:PO}
\end{figure}

{For the HKW cell there exists a library of POs made available by \citet{Gibson2008a}. To investigate if the DManD model finds POs similar to existing solutions, we took states from the known POs, encoded them, and used this as an initial condition in the DManD Newton solver to find POs in the model. In Fig.\ \ref{fig:PO} we show projections of 12 known POs, which we identify by the period $T$, and compare them to projections of POs found using the DManD model.} This makes up a majority of the POs made available by \citet{Gibson2008a}. Of the other known solutions, three are RPOs with phase-shifts in the streamwise direction that our model, with the current setup, can not capture{. The} other two have short periods of $T=19.02$ and $T=19.06$. 
A majority of the POs found with DManD land on initial conditions near that of the DNS and follow similar trajectories trajectories to the DNS. 

How close many of these trajectories are to the true PO is surprising {and encouraging} for many reasons. 
{First, the data used for training the DManD model does not explicitly contain any POs. Second, this approach by no means guarantees convergence on a PO in the DManD model.}
Third, starting with an initial condition from a PO does not necessarily mean that the solution the Newton solver lands on will be the closest PO to that initial condition, so there may exist POs in the DManD model closer to the DNS solutions than what we present here.

Now that we know the DManD model can find POs similar to those known to exist for the DNS, we now use it to search for new POs. First, we searched for POs in three of the $d_h=18$ models by randomly selecting 20 initial conditions and selecting 4 different periods $T=[20,40,60,80]$. We then took the initial conditions and periods for converged periodic orbits and decoded and upsampled them onto a $48 \times 49 \times 48$ grid. We performed this upsampling because \citet{Viswanath2007} reported that solutions on the {coarser} grid can be computational artifacts. Finally, we put these new initial conditions into Channelflow and ran another Newton search for 100 iterations. This procedure resulted in us finding 9 new RPOs {and 3 existing POs}, the details of which we include in Table \ref{Table2}. 

\begin{table}
	\captionsetup{justification=raggedright}
	\caption{Details on the RPOs and POs found using initial conditions from the DManD model. The first 9 solutions are new and the last 3 had previously been found. ``Label" indicates the label in Fig.\ \ref{fig:NewPOb}, $\sigma_z$ corresponds to the phase-shift in $z$, $T$ is the period of the orbit, and ``Error" is $||\text{shifted final state} - \text{initial state}||/||\text{initial state}||$, which is the same error as in \citet{Viswanath2007}.}
	\resizebox{.95\textwidth}{!}{
		\begin{tabular}{r*{14}{c}r}
			Label & \vline & 1 & 2 & 3 & 4 & 5 & 6 & 7 & 8 & 9 & 10 & 11 & 12 \\
			\cline{3-14}
			$\sigma_z$  & \vline & 1.91e-1 & -9.66e-2 & -1.77e-3 & 1.15e-1 & -9.21e-3 & -1.90e-1 & -1.28e-2 & -1.19e-1 & -5.63e-5 & 4.64e-14 & 2.17e-14 & 2.73e-13 \\
			$T$  & \vline & 37.94 & 84.25 & 91.29 & 82.07 & 74.14 & 41.24 & 110.67 & 83.31 & 64.64 & 19.06 & 68.07 & 75.35 \\
			Error & \vline & 2.23e-3 & 1.01e-3 & 3.92e-3 & 2.84e-3 & 1.87e-3 & 5.26e-4 & 1.25e-3 & 1.13e-3 & 2.25e-3 & 1.57e-4 & 2.55e-4 & 1.07e-4
		\label{Table2}
		\end{tabular}}
\end{table}

\begin{figure} 
	\includegraphics[width=\textwidth]{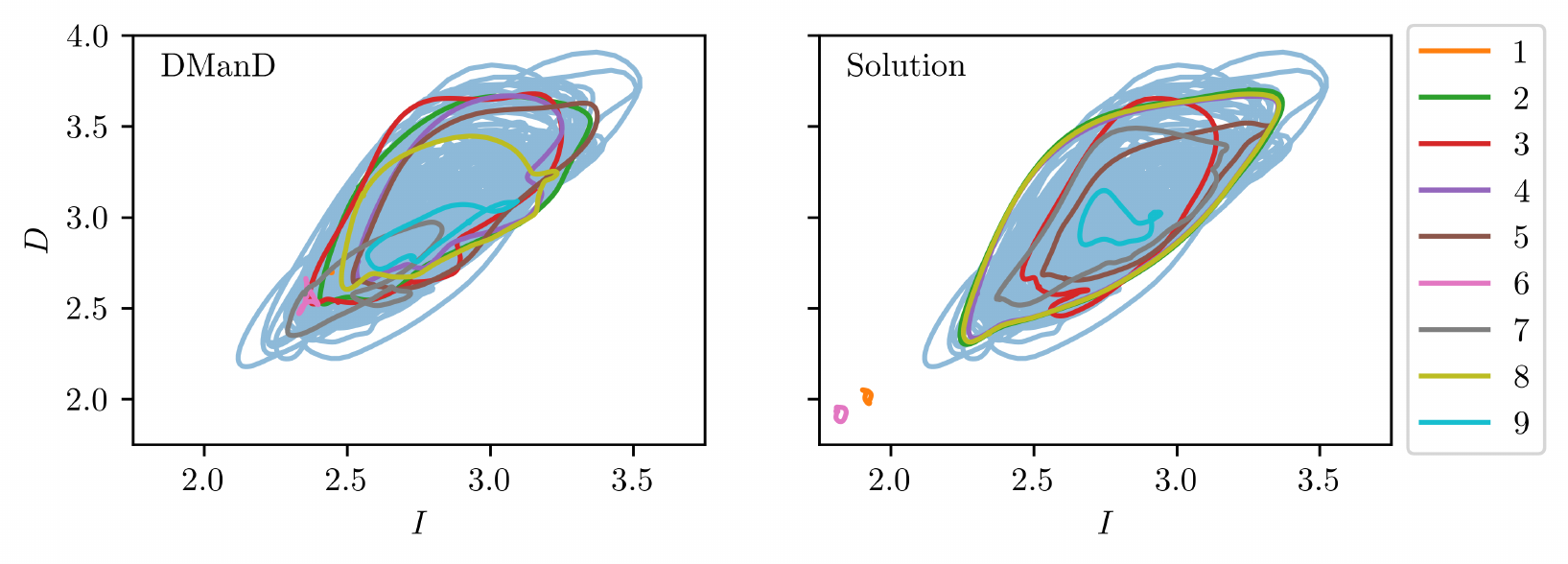}
	\captionsetup[subfigure]{labelformat=empty}
	\begin{picture}(0,0)
	\put(5,130){\contour{white}{ \textcolor{black}{a)}}}
	\put(180,130){\contour{white}{ \textcolor{black}{b)}}}
	\end{picture} 
	\begin{subfigure}[b]{0\textwidth}\caption{}\vspace{-10mm}\label{fig:NewPOa}\end{subfigure}
	\begin{subfigure}[b]{0\textwidth}\caption{}\vspace{-10mm}\label{fig:NewPOb}\end{subfigure}
	
	\caption{(a) Periodic orbits found in DManD models at $d_h=18$ that converged to the (b) periodic orbits found in the DNS. {Each of the colors corresponds to a one of the new solutions in Table \ref{Table2}. The blue curve at the back is a long trajectory of the DNS for comparison.}}
	\label{fig:NewPO}
\end{figure}
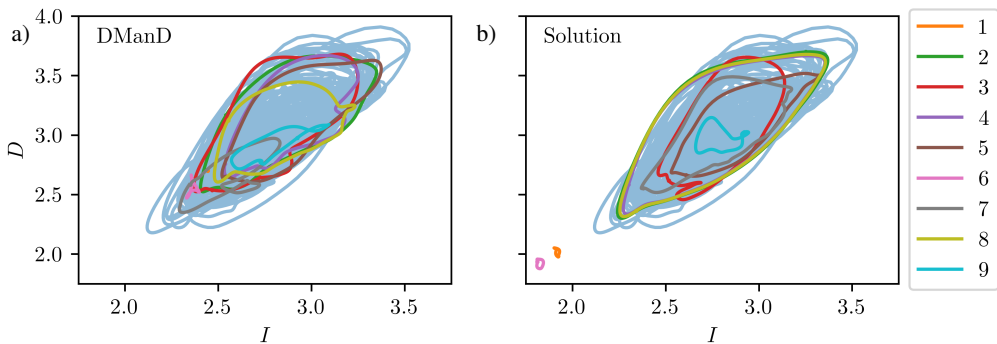

In Fig.\ \ref{fig:NewPOa} we show the new RPOs in the DManD model and in Fig.\ \ref{fig:NewPOb} we show what {they} converged to after putting them into the Channelflow Newton solver {as initial guesses}. Again, many of the RPOs end up following a similar path through this {state} space, with the biggest exceptions being RPO1 and RPO6, which converged to low-power input solutions. It is worth noting that this worked well, considering that the DManD initial conditions are POD coefficients from a model trained using data on a {coarser} grid than used to search for these {solutions. We have described a} new method to rapidly find new ECS, wherein an accurate low-dimensional model, like the DManD model presented here, is used to quickly perform a large number of ECS searches in the model and then these solutions can be fine tuned in the full simulation to land on new solutions. 

\section{Conclusion} \label{sec:Conclusion}

In the present work we {described} a data-driven manifold dynamics method (DManD) and applied it for accurate modeling of MFU PCF with far fewer degrees of freedom ($\mathcal{O}(10)$) than required for the DNS ($\mathcal{O}(10^5)$). The DManD method consists of first finding a low-dimensional parameterization of the manifold on which data lies, and then discovering an ODE to evolve this low-dimensional state representation forward in time. In both cases we use NNs to approximate these functions from data. We find that an extremely low-dimensional parameterization of this manifold can be found using a hybrid autoencoder approach that corrects upon POD coefficients. Then, we use stabilized neural ODEs to accurately evolve the low-dimensional state forward in time.

The DManD model captures the self-sustaining process and accurately tracks trajectories and the temporal autocorrelation over short time horizons. 
For DManD models with $d_h>10$ we found excellent agreement between the model and the DNS in computing the mean-squared POD coefficient amplitude, the Reynolds stress, and the joint PDF of power input vs.\ dissipation. For comparison, we showed that a POD-Galerkin model requires $\sim 2000$ degrees of freedom to get similar performance in matching the mean-squared POD coefficient amplitudes. Finally, we used the DManD model at $d_h=18$ for PO searches. Using a set of existing POs, we successfully landed on nearby POs in the model. Finally, we found 9 previously undiscovered RPOs by first finding solutions in the DManD model and then using those {as initial guesses} to search in the full DNS.

{The results reported here have both fundamental and technological importance. At the fundamental level they indicate that, the true dimension of the dynamics of a turbulent flow can be orders of magnitude smaller than the number of degrees of freedom required for a fully-resolved simulation. Technologically this point is important because it may enable, for example, highly sophisticated model-based nonlinear control algorithms to be used: Determining the control strategy from the low-dimensional DManD model rather than a full-scale DNS, and applying it to the full flow will speed up both learning and implementing a control policy \citep{ZengLinot2022,Zeng.2022.10.1098/rspa.2022.0297}.}
 
\begin{acknowledgments}
This work was supported by AFOSR FA9550-18-1-0174 and ONR N00014-18-1-2865 (Vannevar Bush Faculty Fellowship).
\end{acknowledgments}

\bibliographystyle{jfm}

\end{document}